\documentclass[journal,10pt,twocolumn]{IEEEtran}
\usepackage{amsmath}
\usepackage{subcaption}
\usepackage{mathtools}
\usepackage{mathrsfs}	
\usepackage{amsbsy}
\usepackage{amssymb}
\usepackage[ruled,lined,linesnumbered]{algorithm2e}
\DeclarePairedDelimiter\ceil{\lceil}{\rceil}
\DeclarePairedDelimiter\floor{\lfloor}{\rfloor}
\usepackage{mathtools}
\DeclareMathOperator*{\argmin}{argmin}
\usepackage{tabularx}
\usepackage{algpseudocode}
\usepackage{pifont}
\usepackage{dsfont}
\usepackage{array}
\usepackage{cite}
\usepackage{calc}
\usepackage{color}
\usepackage[dvipsnames]{xcolor}
\usepackage{pdfpages}
\usepackage{graphicx}
\usepackage[normalem]{ulem}
\makeatletter
\renewcommand*\env@matrix[1][*\c@MaxMatrixCols c]{%
  \hskip -\arraycolsep
  \let\@ifnextchar\new@ifnextchar
  \array{#1}}
\makeatother
\newcolumntype{M}[1]{>{\centering\arraybackslash}m{#1}}
\newcolumntype{N}{@{}m{0pt}@{}}
\begin{document}
\title{Ultra-Dense 5G Small Cell Deployment for Fiber and Wireless Backhaul-Aware Infrastructures}

\author{Ali~Lotfi~Rezaabad,
        Hamzeh~Beyranvand,~\IEEEmembership{Member,~IEEE,}
        Jawad A.~Salehi,~\IEEEmembership{Fellow,~IEEE,}
        and~Martin~Maier,~\IEEEmembership{Senior Member,~IEEE}
\thanks{A. Lotfi was with the Electrical Engineering Department, Sharif University of Technology, Tehran, Iran. He is now with Wireless Networking and Communication Group (WNCG), The University of Texas at Austin, USA (e-mail:alotfi@utexas.edu).}
\thanks{H. Beyranvand is with Department of Electrical Engineering, Amirkabir University of Technology, Tehran, Iran (e-mail: beyranvand@aut.ac.ir).}
\thanks{J. A. Salehi is with Department of Electrical Engineering, Sharif University of Technology, Tehran, Iran (e-mail: jasalehi@sharif.edu).}
\thanks{M. Maier is with the Optical Zeitgeist Laboratory, INRS, Montreal, QC, H5A 1K6, Canada (e-mail: maier@ieee.org).}}

\maketitle
\begin{abstract}
In this paper, we study the cell planning problem for a two-tier cellular network containing two types of base stations (BSs)-- i.e. with fiber backhaul, referred to as wired BSs (W-BSs), and BSs with wireless backhaul, referred to as unwired-BSs (U-BSs). In-band full-duplex wireless communications is used to connect U-BSs and W-BSs. We propose an algorithm to determine the minimum number of W-BSs and U-BSs to satisfy given cell and capacity coverage constraints. {\color{black} Furthermore, we apply our proposed non-dominated sorting genetic algorithm II (NSGA-II) to solve both cell planning and joint cell and backhaul planning problem to minimize the cost of planning, while maximizing the coverage simultaneously.} Additionally, the considered cell planning program is developed into an optimization by including the problem of minimizing the cost of fiber backhaul deployment. In order to analyze the performance of the proposed algorithm, we study three different deployment scenarios based on different spatial distributions of users and coverage areas. {\color{black}The results show the superiority of our proposed NSGA-II algorithm for both cell planning and joint cell and backhaul planning to other well-known optimization algorithms.} The results also reveal that there is a trade-off between cell deployment costs and SINR/rate coverage, and W-BSs are placed in congested areas to consume less resources for wireless  backhauls. Similarly, a trade-off between cell and fiber deployment costs and SINR/rate coverage is observed in planning. We show that for realistic scenarios desirable solutions can be selected from the Pareto front of the introduced multi-objective problem based on given cellular operator policies.
\end{abstract}
\begin{IEEEkeywords}
Cell planning, in-band full-duplex (IBFD), fiber-wireless (FiWi) networks, millimeter-wave networks, next-generation passive optical networks (NG-PONs), self-backhauling.
\end{IEEEkeywords}
\IEEEpeerreviewmaketitle

\section{Introduction}

\IEEEPARstart{T}{he ever}-increasing demand for wireless data is insatiable and affecting, immensely, the technology and the design of future wireless networks. In $2015$, mobile data traffic was about $3.7$ Exabytes per month, and based on Cisco's forecasts \cite{Cisco}, it will exceed $30$ Exabytes per month by $2020$. Although wireless communication techniques have been developed to meet this demand, most of them will not suffice to satisfy exponentially increasing mobile data traffic volumes. Recently, numerous studies on 5G cellular networks have been reported. Nonetheless, there remain open questions about the final technologies of choice and 5G standards. According to~\cite{6824752}, 5G will rely on the following three enabling technologies, the so called \textit{big-three}: (i) massive multiple-input multiple-output (MIMO), (ii) millimeter wave (mmWave), and (iii) ultra-densification. Interestingly, these technologies are not only compatible and congruent but also represent prerequisites for each other.

The frequencies between $30$ and $300$ GHz are referred to as mmWave. Due to the high path loss, so far mmWave systems were not at the center of mobile communications research, though undoubtedly they will play a major role in 5G \cite{6736750}. The availability of huge amounts of bandwidth in the mmWave band paves the way to serving mobile users with high data rates \cite{6824746}. Furthermore, the implementation of massive-MIMO in mmWave systems can be easily realized via small-size array antennas.

Telecommunication companies apply densification techniques (i.e., ultra-dense small cell deployment) to improve their areal spectral efficiency, in particular in highly congested regions. Due to the high path loss in the mmWave band densification is inevitable. In \cite{6834753}, it was shown through experimental measurements that the maximum range of mmWave base stations (BSs) is less than $200$ meters. The main problem raised by densification is how to serve all BSs with appropriate backhaul\footnote{ It sould be noted that in the terminology of cloud-radio access network (C-RAN) the word \textit{fronthaul} indicates the link between remote radio head (RRH) and base band unit (BBU), while the term \textit{backhaul} means the backbone infrastructure connecting BBUs to the core network. However, in this study, we use the term \textit{backhaul} to indicate either fiber links between W-BSs and central office or wireless links between U-BSs and W-BSs.}  solutions\cite{6963798}.

\subsection{Related Work}
In\cite{6588652}, existing fiber-to-the-node residential access	networks are utilized to design a cost-optimized optical fiber backhaul for small cells. In \cite{7218503}, the passive optical network (PON) architecture has been utilized as backhaul of heterogeneous networks. In \cite{7306543} and \cite{7115912}, mixed free-space optic (FSO)/radio frequency (RF) and optical fiber backhaul networks were used to serve BSs. It was shown that the integration of optical fiber with other technologies is more practical and cost-effective than leveraging only optical fiber backhaul solutions. More recently, in \cite{7947084}, authors provides an optimization framework to deploy mixed fiber and wireless backhauls for BSs, where the survivability of fiber backhaul has been investigated. In addition, authors in \cite{7925837} proposed a cost effective fiber backhaul deployment and resource allocation optimization for integrated access and backhaul cellular networks where resources are shared dynamically between access and backhaul links.
 
Recently, self-backhauling has been attracting an increasing amount of attention\cite{Interdigital}. On the other hand, given the successful demonstrations of in-band full-duplex (IBFD) radio systems \cite{bharadia2013full}, there is an emerging trend to utilize this technique in self-backhauled wireless systems\cite{7306541}. It is worth mentioning that in IBFD self-backhauling, the same frequency band is simultaneously used for both backhaul and access links.

In \cite{1230131}, the authors investigated the cell planning problem for $3$G cellular networks. They considered a set of candidate locations for BSs and selected optimal locations based on given traffic distributions and power consumption restrictions. In \cite{6196268}, the cell planning problem was investigated from an energy-efficiency point of view by means of ray-trace models. In \cite{7450686}, the authors proposed a cost-optimized cell planning method, whereby the locations of macro- and pico-BSs as well as relay nodes were determined. It is worth mentioning that in \cite{1230131} - \cite{7450686}, a finite set of candidate locations was examined in greater detail. In \cite{7056465}, the authors proposed a cell planning approach by considering infinite candidate locations and various user distributions. Moreover, they utilized meta-heuristic algorithms to solve the cell planning problem under consideration. More recently, in \cite{7883847}, the authors reviewed cell planning problems and also investigated the problem of cell planning for future cellular network.

{\color{black}All the aforementioned studies considered traditional cellular networks using the ultra-high frequency (UHF) band}. In \cite{7305743}, the cell planning problem was resolved for mmWave cellular networks. The authors utilized a meta-heuristic algorithm to obtain near-optimum solutions. Furthermore, in \cite{7450185}, a new approach was proposed for solving the mmWave cell planning problem by leveraging the polygon computational geometry concept. Note, however, that in this study the issue of backhauling was not addressed. {\color{black} There were also some other studies on self-backhauling. Among them is \cite{7386685}, where the authors investigated joint beamforming, power allocation, and spectrum assignment in a self-backhauling two-tier wireless network. Further, cell association and backhaul spectrum assignment were studied in \cite{7505948}. These two studies optimized the aforementioned parameters for fixed BSs, whereas in this work we determine the locations of self-backhauled BSs subject to satisfying the rate and coverage constraints}.

\subsection{Paper Contributions and Organization}
This paper makes the following three main contributions:
\textbf{Self-backhauled mmWave cell planning:}
In this paper, a general optimization model is proposed for mmWave small cell planning. We consider both fiber and wireless backhauling techniques. Accordingly, two types of BSs are deployed, BSs with either fiber or wireless backhaul. At BSs with wireless backhaul, the IBFD technique is used to realize self-backhauled implementation. To satisfy, both, given coverage and capacity constraints with minimum number of either BSs, we propose a multi-objective optimization problem to simultaneously determine number of required BSs with fiber, wireless backhaul, and the optimum locations for BSs. Furthermore, an infinite set of candidate locations is considered.

\textbf{Joint cell and fiber backhaul planning:}
We also formulate another multi-objective optimization problem for joint cell and fiber backhaul planning. The multi-objective problem contains two parts. One part optimizes the cell planning problem and the other one minimizes deployment costs. To reduce the cost of fiber installation, we leverage existing dark fibers and determine the best locations for installing optical splitters. We utilize a meta-heuristic approach to solve this multi-objective problem.

\textbf{Efficient meta-Heuristic algorithm:}
To solve the above problems we develop an efficient meta-heuristic algorithm based on the well-known non-dominated sorting genetic algorithm (NSGA-II). We compare our proposed algorithm to meta-heuristic alternatives utilizing particle swarm optimization (PSO), tabu search  (TS), simulated annealing (SA), and ant colony optimization (ACO). Our simulation results reveal the superiority of the NSGA-II based algorithm to the considered alternatives in terms of the obtained objective and speed of convergence. In addition, by means of simulation, we examine the impact of the probability density function (PDF) used to change the positions of BSs and tune this PDF to improve the objective of both cell planning and joint cell and backhaul planning problem.

{\color{black}It is worth mentioning that all the considered features for the radio access network are in accordance with the 3GPP standards \cite{5Gstandard1} \cite{5Gstandard2}.}

The remainder of the paper is organized as follows. Section II describes the system model. In Section III, the proposed cell planning and joint cell and fiber backhaul planning are formulated. Section IV presents the proposed meta-heuristic algorithms. Numerical results and performance evaluation are presented in Section V. Finally, Section VI concludes the paper.

\section{System Model}

\subsection{Network Architecture}
In our system model, we consider small cells operating in the mmWave band and define two types of small-cell base stations, namely, wired base station (W-BS) and unwired base station (U-BS). As shown in Fig. 1, W-BSs are served with fiber backhaul and U-BSs are connected to W-BSs via wireless backhaul. 

For U-BSs, the wireless backhaul is implemented by means of IBFD, referred to as IBFD wireless self-backhauling \cite{jain2011practical}-\cite{7817893}. With this technique, the same frequency band is simultaneously used for both access and backhaul. The main issue of this technique is the self-interference between access and backhaul links. {\color{black}Hopefully}, this interference can be mitigated efficiently by using separated highly directional antennas, thus providing sufficient distance between transmitter and receiver antennas, and utilizing advanced cancellation techniques in both digital and analog domains to remove residual interference \cite{6832464}. Let  $\tau\times P_a$  denotes the residual self-interference noise in the backhaul link in the downlink direction, where $\tau$ is the percentage of residual self-interference and  $P_a$ is the transmission power of access links.

As illustrated in Fig. 1, we utilize a passive optical network (PON) architecture to realize the fiber backhaul. This architecture has been standardized as a promising scheme to realize fiber-to-the-x (FTTx) deployments, where the x may represent a home, building, neighborhood, or curb \cite{6289432}. The main components of a PON consist of the optical line terminal (OLT), passive power splitter, and optical network units (ONUs). Generally, a PON has a tree-and-branch topology, whereby the OLT and ONUs serve as the root and leave nodes, respectively. The OLT is located at the central office (CO) and performs resource allocation among ONUs, and ONUs reside at subscriber premises. In our model each W-BS is connected to an ONU serving a single PON subscriber. 

In this paper, we focus on the latest standard of PONs, the so-called Next-Generation Passive Optical Network 2 (NG-PON2), which is specified in ITU-T recommendation G.989.2 \cite{ITU}. In NG-PON2, multiple subscribers are served by utilizing hybrid time and wavelength division multiplexing (TWDM), where each subscriber transmits at the maximum data rate of $10$ Gbps \cite{7389581}. By leveraging wavelength division multiplexing (WDM), it is also possible to implement a point-to-point connection between the OLT and any ONU, which is desirable for W-BS backhaul as it provides guaranteed low-latency backhaul with a dedicated capacity of $10$ Gbps. 

Interestingly, NG-PON2 is backward-compatible with previous PON standards such as GPON, which is widely used in FTTx deployments \cite{7389581}. To provide fiber backhaul for small cells, we leverage pre-installed GPON equipment and only upgrade the OLT and splitter. As shown in Fig. 1, conventional GPON subscribers (denoted as  FTTx subscriber) may be served beside W-BSs over the common fiber infrastructure. 

In the proposed joint cell and fiber backhaul planning, by taking existing FTTx  feeder fibers into account we determine the optimal locations for installing new splitters and deploying new distribution fibers. In our approach, the goal is to minimize the cost of new required facilities for fiber backhaul planning. However, we assume there is no available distribution fiber at the optimum place of a W-BS, and a new distribution fiber branch is installed to connect that W-BS to the existing fiber infrastructure.

\begin{figure}[t!]
\centering
\includegraphics[trim={8.5cm 3cm 6.5cm 2.5cm},width=8cm,clip]{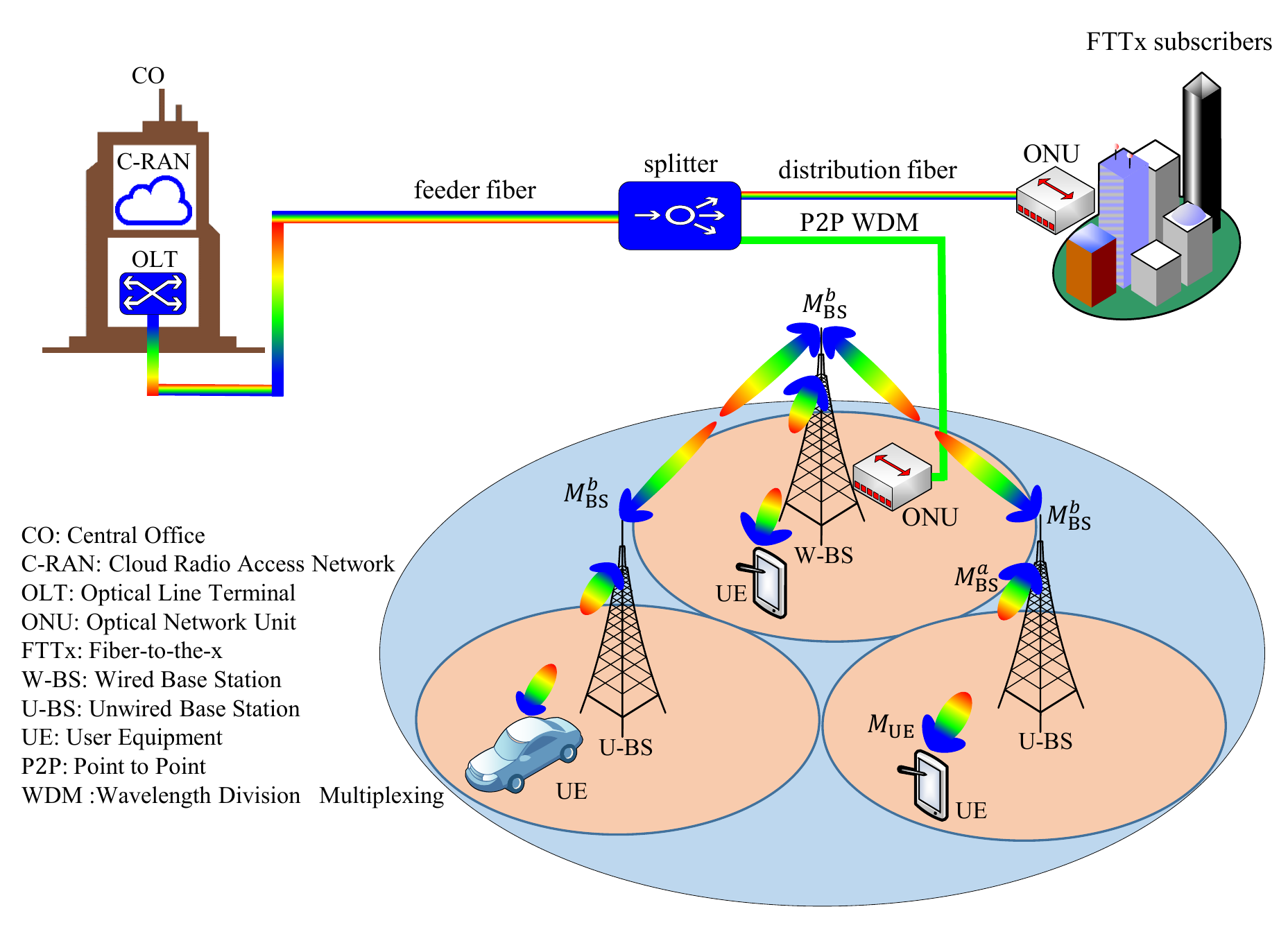}
\caption{Network architecture and its main components.}
\end{figure}

\subsection{mmWave Propagation and Blockage Model}

Two main attributes of cellular networks are capacity and coverage, whereas the most critical design parameter in mmWave networks is path loss. With electromagnetic waves, the increase of path loss exponent is much higher in non line-of-sight (NLOS) than line-of-sight (LOS) systems, especially for higher frequencies such as mmWave \cite{rappaport2013millimeter}. Clearly, in mmWave cellular networks, LOS channels are preferable to increase the spectral efficiency.

The well-known path loss model (in dB) for mmWave cellular network is as follows \cite{singh2015tractable}:
\begin{equation}\label{PathLoss}
\text{PL}(d)=\alpha + 10\beta\log||\pmb{\mathscr{L}}_\text{BS}-\pmb{\mathscr{L}}_\text{UE}||+\zeta,
\end{equation}
where $\alpha$ and $\beta$ denote the path loss at close-in reference distance and the path loss exponent, respectively. Furthermore, we have $\zeta\sim\mathcal{N}(0,\sigma^2)$, where $\sigma^2$ represents the lognormal shadowing variance, $\pmb{\mathscr{L}}_{(.)}$ denotes the position of nodes in Cartesian coordinates, and $||.||$ is the Euclidean distance. Note that $\beta$ as well as $\sigma$ are different for LOS and NLOS.
 
The high path loss in mmWave systems limits the communication range to very short distances. However, this limitation can be partially alleviated by using directional  antennas. In this study, we assume steerable array antennas at BSs for access link and highly directional horn antennas for backhaul links, whereby ${AG}_{\text{BS}}^a$ and ${AG}_{\text{UE}}$ denote the array gain of the access link at BSs and {\color{black} user equipment (UE)}, respectively, and ${AG}_{\text{BS}}^b$ represents the array gain for the backhaul link at BSs. Furthermore, we assume that a perfect array beam matching is performed for both access and backhaul links.

In \cite{6840343}, a stochastic blockage model was derived for mmWave channels based on random shape theory and stochastic geometry. It was shown that when the blockage is modeled as a rectangle Boolean scheme, the probability of the channel status (LOS or NLOS) is given by
\begin{equation}\label{LinkProbability}
\mathbb{P}^{(s)}{(d)}={\begin{cases} 
       e^{-a_{\text{los}}d} & s:\text{LOS} \\
       1-e^{-a_{\text{los}}d} & s:\text{NLOS},
   \end{cases}
}
\end{equation}
where $d$ denotes the link distance and $a_{\text{los}}$ is a parameter that is determined by the density and average size of the blockages in a specific environment. {\color{black}For convenience, the notation and parameters are summarized in Table I.}

\begin{table*}\label{noations}
\caption{{\color{black}Notation and Parameters}}
{\color{black}
\begin{center}
\begin{tabular}{|M{2.1cm}|M{6cm}|M{2.3cm}|M{6cm}|} 
\hline
\centering
\textbf{Notation}& \textbf{Parameter}& \textbf{Notation} & \textbf{Parameter} \\
\hline
$P_a$ & access link transmission power&
$P_b$ & backhaul link transmission power\\
\hline
$\tau$ & residual self-interference percentage for duplex link &
$\alpha$ & path loss at 1 $m$ \\
\hline
$\beta$ & path loss exponent&
$\text{PL}(d)$ & path loss for $d$ $m$\\
\hline
$\sigma^2$ & shadowing variance &
$\textbf{L}_{(\cdot)}$ & location of BSs\\
\hline
$\phi({s_t})$ & users distribution in $s_t$ & $\lambda_{s_t}$ & average of users in $s_t$\\
\hline
$\pmb{\mathscr{L}}_{(\cdot)}$ & position in Cartesian coordinates &
$\mathcal{X}^{(\cdot)}_{(\cdot)}$ and $ ~\mathcal{Y}^{(\cdot)}_{(\cdot)}$ & users and U-BSs association indicator, respectively\\
\hline 
$\Upsilon_{(\cdot)}^{(\cdot)}$& successful detection indicator &
$AG^{(\cdot)}_{(\cdot)}$ & array gain \\
\hline 
$\Gamma^{(\cdot)}_{(\cdot)}$ & SINR & 
$\gamma^{th}_{(\cdot)}$ & SINR threshold for detection\\
\hline
$\varrho^{th}_{(\cdot)}$ & lower bound for successful detection probability & $I_{(\cdot)}$ & interference \\
\hline 
$\sigma_{(\cdot)}$ & large constant for penalty& $d(\mathbf{x})$ & crowding distance\\
\hline
$BW_s$, $BW_{RB}$, and $bw_n$ & available bandwidth for each sector, resource block, and $n$-th user, respectively & $\mathcal{N}$, $\mathcal{P}$,  $\mathcal{B}$, $\mathcal{W}$, $\mathcal{U}$, and $\mathcal{F}$ & sets of users, pixels, BSs, W-BSs, U-BSs, and FAPS, respectively\\
\hline
$\bar{R}_{(DL)}, ~ \text{RB}_{th}$ & average data rate and resource block for each user in the downlink, respectively & $C_W,~C_U$ & costs of implementing W-BS and U-BS, respectively \\
\hline
$FR_i$ & $i$-th Pareto front & $\sigma^2_n$ & thermal noise \\
\hline
$R_n,~r_n$ & achieved and requested data rate for $n$-th user & $F_i(\mathbf{x})$ & $i$-th object\\
\hline 
$\mathbf{F}(\mathbf{x})$ & vector of objects & $N_{\text{U-BS}}$, $N_{\text{W-BS}}$ & number of W-BSs and U-BSs, respectively  \\
\hline
$\delta_{(\cdot)}$ & constraint relaxation & $\mathcal{Z}_f$ & FAP selection \\
\hline
$\mathcal{C}^b_f$ & indicator to determine if $b$-th is connected to $f$-th FAP & $C_s,~C_d,$ and $C_f$ & costs of selecting FAP, distribution, and feeder fibers, respectively\\
\hline
$\Pi^p_i$ & line of sight indicator & $\mathcal{R}$ & FAP capacity\\
\hline
\end{tabular}
\end{center}
}
\end{table*}

\section{Problem Formulation}
\subsection{Assumptions and Preliminary Parameters}
{\color{black} In this paper, we design a cellular network over an area of size $S_T$ divided into  $T$ subareas, $\mathcal{S}_T={\{s_1,s_2,...,s_T}\}$, whereby each subarea  ($s_t$) has a unique user spatial distribution ($\phi({s_t})$) with a different average number of users per unit area ($\lambda_{s_t}$).}

Without loss of generality, we assume that all users request the same average data rate and denote the average data rate in the downlink by $\bar{R}_{(DL)}$. {\color{black} Here, it is important to note that this assumption will be generalized in the next subsection by assuming different data rate demands for each user. With regard to the average data rate, we also assume that at least $\text{RB}_ {th}$  resource blocks should be dedicated to each user to satisfy her demand.} Hence, the largest number of users that can be supported by each BS is obtained as follows:

\begin{equation}\label{maximumnumberofusersperBS}
N^{\text{User}}_{\text{BS}}=\floor{\frac{N_s\times BW_s}{\text{RB}_{th}\times BW_{RB}}},
\end{equation}
where $N_s$ indicates the number of sectors of each BS; $BW_s$ and $BW_{RB}$ denote the available bandwidth of each sector and resource block (RB), respectively; and $\floor{.}$ is the floor function.
The smallest number of BSs to satisfy the users' data traffic demand in each subarea is computed as
\begin{equation}\label{minimumnumberofBSperSubarea}
N^{s_i}_{\text{BS}}=\ceil{\frac{\lambda_{s_i}S_{s_i}}{N^{\text{User}}_{\text{BS}}}},   \qquad\forall s_i\in\mathcal{S},
\end{equation}
where $S_{s_i}$ is the size of subarea $s_{i}$ and $\ceil{.}$ represents the ceiling function. Consequently, the smallest total number of BSs required to serve all data requests in $S_T$ is given by
\begin{equation}\label{numberofBSforCapacity}
N^{\text{Cap}}_{\text{BS}}=\sum_{s_i\in{\mathcal{S}}}N^{s_i}_{\text{BS}}.
\end{equation}
It is worth mentioning that the number of W-BSs has to be equal or larger than $N_{\text{BS}}^{\text{Cap}}$ since all traffic is steered by W-BSs.

Next, the number of required BSs to satisfy the coverage constraint is given by
\begin{equation}\label{coverage}
N^{\text{Cov}}_{\text{BS}}=\ceil{\frac{S_T}{S_{\text{BS}}}},
\end{equation}
where $S_ {\text{BS}}$ is the coverage area of each BS. {\color{black} Then, we initialize our algorithms by feeding  $N_{\text{BS}}=\max\{N^{\text{Cov}}_{\text{BS}},N^{\text{Cap}}_{\text{BS}}\}$ into them. It should be noted that $N_{\text{BS}}$, which is $N_{\text{W-BS}}+ N_{\text{U-BS}}$, is also a variable of our optimization problem. Thus, the final value of $N_{\text{BS}}$  might differ from its initial value}. In fact, by means of these primary calculations the optimization process, will be elaborated upon, can be reduced significantly.  

\subsection{Cell Planning Formulations}
Let the following matrices represent the location of BSs:
\begin{subequations}\label{LocationMatrix}
\begin{align}
\mathbf{L}_{\text{W-BS}}&\triangleq(\pmb{\mathscr{L}}^1_{\text{W-BS}},\pmb{\mathscr{L}}^2_{\text{W-BS}},...,\pmb{\mathscr{L}}^{N_{\text{W-BS}}}_{\text{W-BS}}),\\
\mathbf{L}_{\text{U-BS}}&\triangleq(\pmb{\mathscr{L}}^1_{\text{U-BS}},\pmb{\mathscr{L}}^2_{\text{U-BS}},...,\pmb{\mathscr{L}}^{N_{\text{U-BS}}}_{\text{U-BS}}),\\
\mathbf{L}_{\text{BS}}&\triangleq(\mathbf{L}_{\text{W-BS}},\mathbf{L}_{\text{U-BS}}),
\end{align}
\end{subequations}
where $\pmb{\mathscr{L}}^i_{(.)}=\begin{pmatrix}
x^i_{(.)} \\ y^i_{(.)}
\end{pmatrix}$ indicates the $i$-th BS position in the Cartesian system. For the sake of simplicity and without loss of generality, we assume that all BSs have the same height. Let the binary variable $\mathcal{X}^n_i$ denotes the association state of the $n$-th user to the $i$-th BS, as obtained as follows:
\begin{equation}\label{PixelAssociation}
\mathcal{X}_n^i=\begin{cases}1,& \text{if}\; i=\argmin\limits_{k\in\mathcal{B}}||\pmb{\mathscr{L}}^n-\pmb{\mathscr{L}}^{k}||, \\ 0, & \qquad \qquad \; oth.
\end{cases}
,\; \forall n\in \mathcal{N},
\end{equation}
where $\mathcal{B}$ and $\mathcal{N}$ denote the set of BSs and users, respectively. {\color{black} Indeed, with this variable we can indicate whether the $i$'th BS is the nearest BS to the $n$'th user or not. Using the same definition, we can now define a binary variable $\mathcal{Y}^w_u$ indicating the nearest W-BSs to U-BSs; it is set to $1$ if the $w$'th W-BS is the nearest W-BS to the $u$'th U-BS, and $0$ otherwise}. Moreover, let $\Upsilon^i_n$ indicate if the $n$-th user receives the minimum power from {\color{black}her} associated BS for detection or not, which is given by
\begin{equation}\label{PixelsCoverage}
\Upsilon^{i}_{n}=\begin{cases} 1, & \text{if} \; \mathbb{P}(\Gamma^i_n\geq\gamma^{th}_{a})\times\mathcal{X}^i_n\geq \varrho^{th}_{a}.\\
0, & \qquad \qquad \; oth.
\end{cases}
,\;\forall n\in \mathcal{N},
\end{equation}
where $\Gamma^i_n$ is the received SINR at $n$-th user of the $i$-th BS. Furthermore, $\varrho^{th}_{a}$ denotes the lower bound of the probability and $\gamma^{th}_a$ is the SINR threshold for  {\color{black}the successful} detection. $\Gamma^i_n$ is given by
\begin{equation}\label{SNRForPixelss}
\Gamma^i_n=\frac{\text{P}_a\times AG_{\text{BS}}^{\text{a}}\times {AG}_{\text{UE}}\times \text{PL}^{-1}\color{black}({d_{i,n})}}{\sigma^2_N + I_n},
\end{equation}
{\color{black} where $\text{P}_a$ indicates the transmitted power of the access link. $\text{PL}(d_{i,n})$ shows the path loss as a function of the distance between the $i$-th BS and the $n$-th user. ${AG}_{\text{BS}}^a$ and ${AG}_{\text{UE}}$ denote the array gain of the access link at BSs and UEs, respectively. In addition, $\sigma^2_N$ and $I_n$ denote the thermal noise and the corresponding interference that the $n$-th user suffers from the interfering BSs, respectively.}
Note that \eqref{PixelsCoverage} and \eqref{SNRForPixelss} specify access links. Similarly, we define the following variables for backhaul links:
\begin{equation}\label{UBSCoverage}
\Upsilon^{w}_{u}=\begin{cases} 1 & \text{if} \; \text{P}(\Gamma^{w}_{u}\geq\gamma^{th}_{b})\times\mathcal{Y}^{w}_{u}\geq \varrho^{th}_{b}\\
0 & \qquad \qquad \; oth.
\end{cases}
,\;\forall u\in \mathcal{U},
\end{equation}

\begin{equation}\label{SNRbackhual}
\Gamma^{w}_{u}=\frac{\text{P}_b\times ({AG}_{\text{BS}}^b)^2\times \text{PL}^{-1}(d_{u,w})}{\sigma^2_N+\tau\times\text{P}_a+I_u}
\end{equation}
{\color{black} where in \eqref{UBSCoverage} $\mathcal{U}$ denotes the set of U-BSs,  $\gamma^{th}_b$ is the SINR threshold for successful detection, and $\varrho^{th}_b$ is the lower bound of the successful detection probability. In \eqref{SNRbackhual}, $p_b$ denotes the transmission power of the backhaul link, $\text{PL}(d_{u,w})$ is the path loss between two BSs at distance $d_{u,w}$, and  ${AG}_{\text{BS}}^b$ represents the array gain of the backhaul link. Further, we use $\tau\times \text{P}_a$ and $I_u$ to represent the self-interference from the access link and neighboring U-BSs, respectively.}

The primary goal of our cell planing framework is to maximize the number of users served with their requested data rates, $F_1$, which is accepted as crucial aim for 5G. Accordingly, we set the objective function of the cell planning problem to minimize the total number of users which are not satisfied with their demanded rates. 
\begin{equation}\label{indicator}
F_1 = \sum_{n\in\mathcal{N}}I\{R_n<r_n\},
\end{equation}
{\color{black} where $r_n$ and $R_n$ indicate the desired and actual data rate of the $n$-th user, respectively. The indicator function in \eqref{indicator}, $I\{\cdot\}$, equals $1$ if the condition is satisfied, and $0$ otherwise. Here, $r_n$ is a given constant and $R_n$ is given by}

\begin{equation}\label{F1}
R_n = \quad  bw_n \log_2(1+\Gamma^i_n),
\end{equation}
where $bw_n$ is the bandwidth dedicated to the $n$-th user from its associated BS.

It is worth mentioning that another critical aspect of cell planning is to minimize capital expenditure (CAPEX). Hereupon, in addition to the cellular network rate, implementation {\color{black} cost} is also involved in our cell planning procedure. Obviously, the major CAPEX of cellular network has its roots in BSs implementation and feeding them with appropriate backhauls. Therefore, we model the major cost of cell planning $F_2$ as follows:

\begin{equation}\label{F2}
F_2 = N_\text{W-BS}C_W + N_\text{U-BS}C_U,
\end{equation}
where $C_W$ and $C_U$ are indicating the costs of implementing W-BS and U-BS and feeding them with suitable backhaul, respectively. Given the aforementioned definitions and assumptions, our cell planning problem can be formulated as follows:
\begin{alignat}{2}
\label{CellObject}
\underset{\mathbf{L}_\text{W-BS},\mathbf{L}_\text{U-BS},N_\text{W-BS},N_\text{U-BS}}{\text{\textbf{minimize}}}
& [F_1, F_2] 
&&\\ \label{pixelcoverage}
\text{\textbf{subject to:}} \nonumber\\
&\sum_{\forall b\in\mathcal{B}}\sum_{\forall p\in\mathcal{P}}\Upsilon^b_p\geq\delta_{cov}|\mathcal{P}|,
&&\\ \label{UBScoverage}
&\sum_{\forall b\in\mathcal{W}}\Upsilon^{b}_{b'}=1,
~~~~~~~~~~~~\forall b'\in\mathcal{U},\\ \label{limitation}
&\sum_{\forall b'\in\mathcal{U}}\Upsilon^{b}_{b'}\leq N_{lim},
~~~~~~~\forall b\in\mathcal{W},\\ \label{UBSCapacity}
&\sum_{\forall n\in\mathcal{N}}\Upsilon^b_n bw_n\leq N_s BW_s,
~ \forall b\in\mathcal{U},\\ \label{WBSCapacity}
\nonumber
& \sum_{\forall n\in\mathcal{N}}\sum_{\forall b'\in\mathcal{U}}\Upsilon^{b'}_{n}\Upsilon^{b}_{b'} bw_n
&& \\
+
& \sum_{\forall n\in\mathcal{N}}\Upsilon^b_n bw_n\leq N_s BW_s
~ \forall b\in\mathcal{W},\\ \label{CapacityCoverage}
& \sum_{\forall n\in\mathcal{N}}\sum_{\forall b\in\mathcal{B}}\Upsilon^b_n\geq\delta_{cap} |\mathcal{N}|
&&
\end{alignat}

{\color{black}\textbf{Objects}: The optimization problem has two objectives, $F_1$ and $F_2$. Thus, we have a multi-objective problem here. Obviously, these two objects are conflicting since having a good coverage implies having more BSs, which is costly.}
 
\textbf{Coverage constraints}: Constraint \eqref{pixelcoverage} insures the coverage of deployed cells, where $|\cdot|$ denotes the total number of elements in a set, and $\delta_{cov}$ is a constant, relaxing the constraint. Furthermore, we divide the entire region into enough small areas called as pixels, in which $\mathcal{P}$ and $\Upsilon^{b}_{p}$ are denoting the pixel set and SINR at the middle of the $p$-th pixel, respectively. Note that $\Upsilon^{b}_{p}$ is computed using \eqref{PixelsCoverage} for a user located at the center of $p$-th pixel. Constraint \eqref{UBScoverage} is included to associate U-BSs to W-BSs. It is worth mentioning that in our cell planning each U-BS is served by a single W-BS.

\textbf{Limitations of W-BSs}:
Due to the limited fiber backhaul capacity, energy constraint, and computation processing, the total number of U-BSs served by each W-BS is less than $N_{lim}$, as stated in \eqref{limitation}.

\textbf{Capacity Constraints}:
Clearly, the total spectrum of U-BSs assigned to users has to be less than its capacity, which is given by \eqref{UBSCapacity}. The same approach can be employed for W-BSs, while accounting for their distinctive feature of supporting both access and backhaul links simultaneously. 

Two different approaches can be applied to allocate the bandwidth of W-BSs to access and backhaul links: static or dynamic allocation. In static allocation, the bandwidth of each W-BS is divided into two sub-bands, which are separately dedicated to access and backhaul links. Conversely, in dynamic allocation, the W-BS bandwidth is divided among access and backhaul links dynamically according to current traffic loads and bandwidth requests. Clearly, dynamic allocation enables more flexibility and efficiency. It is worth mentioning that this functionality can be easily implemented in cellular networks thanks to the advent of software defined wireless network (SDWN)\cite{6994333} as well as cloud radio access network (C-RAN)\cite{6898939}. Constraint \eqref{WBSCapacity} guarantees that the total bandwidth of a given W-BS allocated to its associated access and backhaul links is less than its available bandwidth.
In addition, constraint \eqref{CapacityCoverage} is included to satisfy an acceptable amount of users, in which, $\delta_{cap}$ is a relaxing constant (similar to $\delta_{cov}$ in \eqref{pixelcoverage}).

\subsection{Joint Cell and Fiber Backhaul Planning}
In this subsection, we optimize both the cell planning and fiber backhaul designing problems. We assume that there are some pre-deployed fiber access points (FAPs) that are randomly distributed across the area. Furthermore, we assume that there are enough dark feeder fibers associated with these FAPs. Our goal is to leverage these dark feeder fibers and FAPs to implement NG-PON2 architectures as the backhaul of W-BSs. 

Let the binary variable $\mathcal{Z}_f$ indicates the selection state of FAPs, which is $1$ if the $f$-th FAP is selected, and $0$ otherwise. In addition, we define a binary variable $\mathcal{C}^{b}_{f}$, which is set to $1$ if the $b$-th W-BS is connected to the $f$-th FAP; otherwise it is set to $0$.

To formulate the cost of the fiber backhaul in our planning problem, we define some parameters in the following. Let $C_s$ indicate the cost of selecting an FAP and installing an optical splitter at it. The costs to utilize feeder fiber and deploy distribution fibers per unit length are denoted by $C_f$ and $C_d$, respectively. Furthermore, let $\ell^{f}_{b}$ indicates the distance between the $f$-th {\color{black}FAP} and the $b$-th W-BS. Similarly, let $\ell^{C}_{f}$ represents the distance between the central office and the $f$-th FAP. 

Given the aforementioned parameters, the total cost of deploying fiber backhaul based on an NG-PON2 architecture ($F_3$) can be modeled as follows:
\begin{equation}\label{F3}
F_3=\sum_{\forall b\in\mathcal{W}}\sum_{\forall f\in\mathcal{F}}\mathcal{C}^b_f C_d\ell^{f}_{b}+\sum_{\forall f\in\mathcal{F}}\mathcal{Z}_f(C_s+C_f\ell^{C}_{f}).
\end{equation}
We apply the same procedure utilized in the previous subsection for cell planning. 

Finally, we can summarize the multi-objective joint cell and fiber backhaul planning as follows:
\begin{alignat}{2}
\label{MOP}
\underset{\mathbf{L}_\text{W-BS},\mathbf{L}_\text{U-BS},N_\text{W-BS},N_\text{U-BS},\mathcal{C}^b_f,\pmb{\mathcal{Z}}}{\text{\textbf{minimize}}}
& \,\,\, [F_1, F_2 , F_3]
&&\\ \label{WBSconnection}
\!\!\!\!\!\textbf{subject to:}
& \sum_{\forall f\in\mathcal{F}}\mathcal{C}^b_f=1,
&& \! \!\!\!\!\!\!\!\!\!\!\!\!\!\forall b \in \mathcal{W},\\ \label{Association}
&\mathcal{C}_f^b\leq\mathcal{Z}_f,
&&\!\!\!\!\!\!\!\!\!\!\!\!\!\forall b\in\mathcal{W},\\ \label{Ratio}
&\sum_{\forall b\in \mathcal{W}}\mathcal{C}^b_f\leq\mathcal{R},
&&\!\!\!\!\!\!\!\!\!\!\!\!\!\forall f\in \mathcal{F},\\
\nonumber
&\text{Constraints~} (\ref{pixelcoverage})-(\ref{CapacityCoverage}),
&&
\end{alignat}
where $F_1$, $F_2$, and $F_3$ are given in \eqref{F1}, \eqref{F2}, and \eqref{F3}, respectively. Also, constraint \eqref{WBSconnection} ensures that only one splitter is associated with each W-BS. Constraint \eqref{Association} states that each W-BS is connected to a selected FAP and constraint \eqref{Ratio} accounts for the capacity limitations of the fiber backhaul, which guarantees that the total number of W-BSs connected to a specific FAP does not exceed the fiber capacity allocated to that FAP ($\mathcal{R}$). 

{\color{black} It should be noted that both $F_2$ and $F_3$ are the deployment costs of wireless and fiber networks. Therefore, they can be merged and treated as one object, $F'_2$, defined as $F_2+F_3$. Thus, the planning objective can be rewritten as $[F_1, F'_2]$}

\section{Proposed Cell Planning Algorithms}
Generally, the cell planning problem, similar to the well-known facility location problem \cite{drezner2001facility}, is NP-Hard \cite{1230131}. Thus, obtaining its optimal solution for large areas with infinite BS candidate positions is not straightforward. Alternatively, meta-heuristic algorithms such as tabu search, genetic algorithms, simulated annealing, and ant colony optimization can be used. The meta-heuristic approaches for cell planning problem can be divided in two categories based on either finite or infinite candidate locations for BSs. Clearly, considering infinite candidate locations results in a better performance at the expense of higher computational complexity. 

In this paper, we assume infinite candidates and utilize genetic algorithm to solve our cell planning problem. Furthermore, we use NSGA-II to solve our planning problems to approach promising solutions. In what follow, first we start with introducing genetic algorithm then we follow NSGA-II algorithm.   

\subsection{Cell planning with meta-heuristic algorithms}
\subsubsection{Genetic Algorithm}
The genetic algorithm (GA) starts with an initial random population (individuals) denoted by $\text{L}^{(n)}_{\text{BS}}, n\in{\{1,2,...,N}\}$, where $N$ is the total number of individuals. 

Let $N_\text{It}$ denotes the total number of iterations in GA. Let define $\mathcal{N}_c$ and $\mathcal{N}_m$ denoting the sets of selected individuals for crossover and mutation, respectively. Note that $|\mathcal{N}_c|$ and $|\mathcal{N}_m|$ are proportional to the total population $N$. In each iteration, $n_c$ individuals are selected from the set $\mathcal{N}_c$ and $n_m$ individuals are selected from the set $\mathcal{N}_m$  for crossover and mutation, respectively. Different strategies may be applied for the selection step, either purely random selection or considering the cost function, for more information we refer interested readers to \cite{996017}.

{\color{black}Let ${\upsilon_1},\,{\upsilon_2}\in \mathcal{N}_c$ denote the selected individuals for crossover. Then, in each iteration the crossover procedure is performed according to Eq. \eqref{RealCrossOver}, and new individuals, $c_1$ and $c_2$, are derived as follows:}
\begin{subequations}\label{RealCrossOver}
\begin{align}
\mathbf{L}^{(c_1)}_{\text{BS}}&=\mathbf{M}\circ\mathbf{L'}^{(\upsilon_1)}_{\text{BS}}+\bar{\mathbf{M}}\circ\mathbf{L'}^{(\upsilon_2)}_{\text{BS}},\\
\mathbf{L}^{(c_2)}_{\text{BS}}&=\bar{\mathbf{M}}\circ\mathbf{L'}^{(\upsilon_1)}_{\text{BS}}+\mathbf{M}\circ\mathbf{L'}^{(\upsilon_2)}_{\text{BS}},
\end{align}
\end{subequations}
where $\circ$ is the sign of the Hadamard production and  $\mathbf{L'}^{(\cdot)}_{(\cdot)}$ is directly derived from $\mathbf{L}^{(\cdot)}_{(\cdot)}$ to compute the summations represented in \eqref{RealCrossOver}. It should be noted that $\upsilon_1$ and $\upsilon_2$ may have different number of elements, thus, the minimum dimension from the pair of $\upsilon_1$ and $\upsilon_2$, or even the maximum dimension can be selected, where for the case of maximum selection random elements are added to which has less elements. Additionally, $\mathbf{M}$ is a random matrix defined as follows:
\begin{equation}\label{RandM}
\begin{aligned}
&\mathbf{M}=\\
  &\begin{pmatrix}[cccc|cccc]
   \mu^{\text{W-BS}}_{x_1} & \!\! \mu^{\text{W-BS}}_{x_2} &\!\! \dots &\!\! \mu^{\text{W-BS}}_{x_{k_1}} & \mu^{\text{U-BS}}_{x_1} &\!\! \mu^{\text{U-BS}}_{x_2} & \!\! \dots & \!\!\mu^{\text{U-BS}}_{x_{k_2}}\\
   \mu^{\text{W-BS}}_{y_1} & \!\!\mu^{\text{W-BS}}_{y_2} & \!\!\dots & \!\!\mu^{\text{W-BS}}_{y_{k_1}} & \mu^{\text{U-BS}}_{y_1} &\!\! \mu^{\text{U-BS}}_{y_2} &\!\! \dots &\!\! \mu^{\text{U-BS}}_{y_{k_2}} \\
  \end{pmatrix},
\end{aligned}
\end{equation}
in which $k_1$ and $k_2$ are selected based on one of the above approaches. In addition, we have $\bar{\mathbf{M}}= \mathds{1}-\mathbf{M},$ where $\mathds{1}$ is an all-one matrix with the same dimension as $\mathbf{M}$. Moreover, we determine the distribution of $\mu^{(\cdot)}_{{(\cdot)}_{(\cdot)}}$ in \eqref{RandM} empirically from simulation results as follows:
\begin{equation}
f(\mu^{(\cdot)}_{(\cdot)_{(i)}})=
\begin{cases}\frac{2}{3(2\epsilon+1)},&\quad -\epsilon\leq\mu^{(.)}_{(\cdot)_{(i)}}\leq 1+\epsilon,\\ 
\frac{1}{3},&\!\!\!\!\!\!\!\!\!\!\!\!\mu^{(\text{W-BS})}_{(\cdot)_{(i)}}=1, \forall i;\,\, -\epsilon\leq\mu^{(\text{U-BS})}_{(\cdot)_{(i)}}\leq 1+\epsilon,\\
\frac{1}{3},&\!\!\!\!\!\!\!\!\!\!\!\!\mu^{(\text{U-BS})}_{(\cdot)_{(i)}}=1, \forall i;\,\, -\epsilon\leq\mu^{(\text{U-BS})}_{(\cdot)_{(i)}}\leq 1+\epsilon,\\
\end{cases}
\end{equation}
where $\epsilon$ is a small positive constant, enabling the search of more possible solutions. {\color{black} Based on $f(\mu)$, we have the following three different possibilities in each iteration: 1) transferring the locations of W-BSs to the next iteration while merging the locations of U-BSs, 2) transferring the locations of U-BSs to the next iteration while merging the locations of W-BSs, and 3) merging the locations of both W-BSs and U-BSs. Using this randomness implied by $f(\mu)$ will help us keep the positions of those BSs that are appropriately located while merging the other. With this policy, we can significantly reduce the number of iterations and improve the performance of the optimization. Further details will be provided in our results below. Note that in the ordinary NSGA-II we just have the third case, where we merge the locations of all BSs.} Let assume $\mathbf{L}^i$ ($i\in\mathcal{N}_m$) is selected for permutation, then, its $j$-th column ($j$ is randomly selected within $[1 ,N_{\text{BS}}]$) is mutated as follows:
\begin{subequations}\label{RealMutation}
    \begin{align}
      \qquad\pmb{\mathscr{L}}^{j}_{\text{BS}}&=\pmb{\mathscr{L}}_{\text{BS}}^{j}+\begin{pmatrix}
      \alpha_x\\
      \alpha_y
      \end{pmatrix}\\
      \sigma_x&=\mu({x_{\text{max}}-x_{\text{min}}}),\\
      \sigma_y&=\mu({y_{\text{max}}-y_{\text{min}}})
    \end{align}
\end{subequations} 
where $\alpha_x\sim\mathcal{N}(0,\sigma_x)$ and $\alpha_y\sim\mathcal{N}(0,\sigma_y)$ are zero mean Gaussian random variables with variances of $\sigma_x$ and $\sigma_y$, respectively. Furthermore, $\mu$ is a small enough positive constant denoting the mutation rate, and $x$ and $y$ denoting Cartesian coordinates. It should be noted that subscripts max and min are set to make sure we have most of the mutated individuals in the allowed area.

To complete the proposed GA, the following penalty functions are included in the objective function of the planning problem, in which we categorize \eqref{pixelcoverage}, \eqref{limitation}, \eqref{UBSCapacity}, and \eqref{CapacityCoverage} as  inequality constraints ($C^{(I)}_j(\mathbf{x})$),  and \eqref{UBScoverage} as equality constraint ($C^{(E)}_j(\mathbf{x})$):{\color{black}
\begin{subequations}
    \begin{align}
      &C^{(I)}_j(\mathbf{x})\geq C_j,\quad\forall j\in{\{1,2,...,N_{C_I}}\}, \\
      &P^{(I)}_j(\mathbf{x}) = \sigma^{(I)}_{v_j}\times\max(1-\frac{C^{(I)}_j(\mathbf{x})}{C_j},0),\\
      &C^{(E)}_j(\mathbf{x})= C_j,\quad\forall j\in{\{1,2,...,N_{C_E}}\},\\
      &P^{(E)}_j(\mathbf{x}) = \sigma^{(E)}_{v_j}\times|1-\frac{C^{(E)}_j(\mathbf{x})}{C_j}|,
    \end{align}
\end{subequations}}
{\color{black}where $\mathbf{x}$ is a vector denoting the optimization variables.} Furthermore, $N_{C_I}$ and $N_{C_E}$ are the number of inequality and equality constraints, respectively; $\sigma^{(I)}_v$ and $\sigma^{(E)}_v$ are large enough positive violation coefficients for inequality and equality constraints, respectively. {\color{black} For instance, the penalty function for \eqref{pixelcoverage} is:
\begin{equation}
\begin{aligned}\label{pentaly example}
& P_1(\mathbf{L}_\text{W-BS},\mathbf{L}_\text{U-BS},N_\text{W-BS},N_\text{U-BS}) =\\ &\sigma_{v_1}\times\max(1-\frac{\sum_{\forall b\in\mathcal{B}}\sum_{\forall p\in\mathcal{P}}\Upsilon^b_p}{\delta_{cov}|\mathcal{P}|},0).
\end{aligned}
\end{equation}
It should be noted that $\Upsilon^{b}_p$ is a function of $\mathbf{L}_\text{W-BS},~\mathbf{L}_\text{U-BS},~N_\text{W-BS},$ and $N_\text{U-BS}$. Finally, using the aforementioned penalty functions, the penalized objectives can be expressed as: 
\begin{equation}
F_i(\mathbf{x})=F_{i}(\mathbf{x})+\sum_{j=1}^{N_{C_I}}P^{(I)}_j(\mathbf{x})+\sum_{j=1}^{N_{C_E}}P_j^{(E)}(\mathbf{x}), \quad \forall i.
\end{equation}
Here, we use $\mathbf{x}$ as a variable vector to abbreviate the variables $\mathbf{L}_\text{W-BS},~\mathbf{L}_\text{U-BS},~N_\text{W-BS},$ and $N_\text{U-BS}$. Since the penalized objective equals the objective itself in the feasible set we use the same notation for them.}

\subsection{Meta-heuristic algorithms for the proposed multi objective problems}
We utilize the well-known  NSGA-II algorithm \cite{srinivas1994muiltiobjective} to solve the considered multi-objective problems. {\color{black}In so doing, Algorithm 1 present GA procedure, afterward, we proceed to NSGA-II for our proposed planning problems.} 

\begin{algorithm}
\SetKwInOut{Input}{Input}\SetKwInOut{Output}{Output}
\caption{Genetic algorithm}
\Input{$N_{\text{It}},\mathcal{N},{\color{black}N_{\text{BS}}},f(\mu^{(.)}_{(i)}),\epsilon,\sigma_\mu,x_{\text{max}}, x_{\text{min}},y_{\text{max}}, y_{\text{min}}$}
\Output{$\mathbf{L}_{\text{W-BS}},\mathbf{L}_{\text{U-BS}},\mathbf{L}_{\text{BS}}$}
{Generate an initial random population containing $N$ chromosomes $\mathbf{L}^{(n)}_{\text{BS}}$, $n=1,2,\dots,N$ and compute their corresponding costs.\\ $F={\{F(\mathbf{x}_1),F(\mathbf{x}_2),\dots,F(\mathbf{x}_N)}\},$}

\For{$t=1,2,\dots,N_{\text{It}}$}{\For{$i=1,2,\dots,{n_c}/{2}$}{Select two individuals from $\mathcal{N}_c$, combine them based on \eqref{RealCrossOver}, and  
compute newborns' costs $F_{c_1}$ and $F(\mathbf{x}{c_2})$.\\ $F\leftarrow\{F,\{F(\mathbf{x}_{c_1})\},\{F(\mathbf{x}_{c_2})\}\}$}\For{$j=1,2,\dots,n_m$}{Select an individual, mutate it based on \eqref{RealMutation}, and calculate its cost $F_m$.\\ $F\leftarrow{\{F,\{F(\mathbf{x}_m)}\}\}$}}
\end{algorithm}

Generally, multi-objective optimization problems have more than one optimal solution. The set of these optimal solutions is referred to as the \textit{Pareto front}. The Pareto front is defined as follows:

{\color{black}\textbf{Definition 1:} \textit{A solution vector $\mathbf{F}(\mathbf{x}^*)=[F_1(\mathbf{x^*}), F_2(\mathbf{x^*}),\dots,F_K(\mathbf{x^*})]$, where $F_i(\mathbf{x})$ indicates the $i$-th objective function, is a Pareto front or non-dominated, if and only if there is no $\mathbf{F}(\mathbf{x})$ such that:
\newline 
$1$) $\forall i, \quad F_i(\mathbf{x})\leq F_i(\mathbf{x}^*)$ and\newline $2$) $\exists i, \quad F_i(\mathbf{x})<F_i(\mathbf{x}^*)$.\\
Similarly, $\mathbf{x}$ dominates $\mathbf{y}$ ($\mathbf{x} \prec_d \mathbf{y}$), if and only if:\\
$1$) $\forall~i \qquad x_i\leq y_i$, and,\\
$2$) $\exists~i_0 \qquad x_{i_0}\leq y_{i_0}.$}}

Two approaches are utilized to find the optimal Pareto front set. The first one is based on decomposition, in which the multi-objective problem is formulated as an one-objective optimization problem. The $\epsilon$-constraint, weighted sum, and goal attainment methods belong to this category \cite{996017}. In this approach, however, multiple executions of these algorithms with various conditions is challenging. The second approach deals directly with the multi-objective optimization problem. In our study, we apply NSGA-II, which belongs to the second category, as explained in more detail next.

\textbf{Non-Dominated Sorting Genetic Algorithm II:}
This algorithm was introduced in \cite{996017}. It leverages GA for evolutionary computing and a specific sorting method referred to as non-dominated sorting. The sorting method is used to rank the solutions of each iteration. In what follows, the parameters required to develop the proposed NSGA-II algorithm are explained.
\begin{algorithm}
\SetKwInOut{Input}{Input}\SetKwInOut{Output}{Output}
\Input{$\mathbf{F}(\mathbf{x}_1),\mathbf{F}(\mathbf{x}_2),\dots,\mathbf{F}(\mathbf{x}_N)$}
\Output{$FR_i$}
\For{$p=1,2,\dots,N$}{\For{$q=p+1,p+2,\dots,N$}{\uIf{$\mathbf{F}(\mathbf{x}_{p})\prec_d\mathbf{F}(\mathbf{x}_q)$}{$\mathcal{S}_p\gets{\{\mathcal{S}_p,\mathbf{F}(\mathbf{x}_q)}\}, n_q\gets n_q+1.$}\uElseIf{$\mathbf{F}(\mathbf{x}_q)\prec_d\mathbf{F}(\mathbf{x}_p)$}{$\mathcal{S}_q\gets\{\mathcal{S}_q,\mathbf{F}(\mathbf{x}_p)\}, n_{p}\gets n_{p}+1$.}{\textbf{end}}}}
\For{$i=1,2,\dots,N$}{$FR_1\gets{\{FR_1,\{\mathbf{F}(\mathbf{x}_i)}\}\},$ \textbf{if} $n_i=0,$}
$k'\gets 1$

\While{true}{$Q\gets{\{}\}$

\For{$\forall \mathbf{F}(\mathbf{x}_i)\in FR_k$}{\For{$\forall \mathbf{F}(\mathbf{x}_j)\in \mathcal{S}_i$}{$n_j\gets n_j-1$

\If{$n_j=0$}{$Q\gets{\{Q,\{\mathbf{F}(\mathbf{x}_j)}\}\}$}}
}

\If{$Q={\{}\}$}{\textbf{break}}
$FR_{k'}\gets Q$; $k'\gets k'+1$}
\caption{Non-dominated sorting}
\label{NS}
\end{algorithm}

Let $\mathcal{S}_{\upsilon}$ indicate a set for $\upsilon$-th individual, which is fulfilled for those individuals dominated by the $\upsilon$-th individual. In addition, a scalar $n_{\upsilon}$ is used to denote the time instances that the $\upsilon$-th individual is dominated by other individuals. The non-dominated sorting procedure is explained in Algorithm \ref{NS}.
{\color{black}Note that in Algorithm \ref{NS}, $\mathbf{F}(\mathbf{x}_j) = [F_{1}(\mathbf{x}_j),~F_{2}(\mathbf{x}_j),\dots,F_{K}(\mathbf{x}_j)]$, where $FR_i$ indicates the $i$-th Pareto front set, i.e., $FR_1$ is a set of solution vectors, $\mathbf{F}(\mathbf{x})$, which are dominated neither by themselves nor by the other solution vectors.}

{\color{black}Ranking the solutions that are far apart from each other by using a prerogative approach is a promising approach to obtain a Pareto optimal solution. To implement this ranking method, we define the so-called crowding distance for the $j$-th answer $d(\mathbf{x}_j)$ as follows:
\begin{equation}\label{CrowdingDistance}
\begin{aligned}
d(\mathbf{x}_j)&=\sum_{i=1}^{K}|\frac{{F_{i}(\mathbf{x}_{j+1})-F_{i}(\mathbf{x}_{j-1})}}{F_i(\mathbf{x}_{\max})-F_i(\mathbf{x}_{\min})}|,
\end{aligned}
\end{equation}
where $\mathbf{F}(\mathbf{x}_{j-1})$ and $\mathbf{F}(\mathbf{x}_{j+1})$ are the two nearest solutions to $\mathbf{F}(\mathbf{x}_j)$ from different sides and are in the same Pareto front as $\mathbf{F}(\mathbf{x}_j)$. Also, $\mathbf{F}(\mathbf{x}_{\max})$ and $\mathbf{F}(\mathbf{x}_{\min})$ are the two head solutions in each front. Note that the crowding distance for these two head solutions needs to be initialized with $\infty$ to avoid being omitted or ignored during this procedure. Solutions with a small crowding distance are likely to be excluded from the procedure in subsequent iterations.}

The variables of the fiber backhaul planning problem are binary. Thus, the crossover and mutation procedures are obtained by modifying Eqs. \eqref{RealCrossOver} and \eqref{RealMutation}. Assume that $\pmb{\mathcal{Z}}^{(\upsilon_1)}$ and $\pmb{\mathcal{Z}}^{(\upsilon_2)},\quad \upsilon_1,\upsilon_1\in \mathcal{N}_c,$ are selected as parents. The binary crossover is then given by
\begin{equation}\label{binarycrossover}
\pmb{\mathcal{Z}}^{(c_1)}=\pmb{\alpha}\circ\pmb{\mathcal{Z}}^{(\upsilon_1)}+\bar{\pmb{\alpha}}\circ\pmb{\mathcal{Z}}^{(\upsilon_2)},
\end{equation}
where $\pmb{\mathcal{Z}}^{(.)}=(\mathcal{Z}^{(\cdot)}_1,\mathcal{Z}^{(\cdot)}_2,\dots,\mathcal{Z}^{(\cdot)}_{N_\mathcal{F}})$, $N_\mathcal{F}$ is the total number of FAPs, $\boldsymbol{\alpha}$ is a $1\times N_{\mathcal{F}}$ binary random vector whose elements are $0$ or $1$ with equal probability, and $\bar{\pmb{\alpha}}=\mathds{1}-\pmb{\alpha}$. Furthermore, the binary mutation for $m\in\mathcal{N}_m$ is given by
\begin{equation}\label{binarymutation}
\mathcal{Z}^m_i\leftarrow 1-\mathcal{Z}^m_i,\quad 1\leq i\leq N_{\mathcal{F}},
\end{equation}
where $i$ is a random integer number between $1$ and $N_{\mathcal{F}}$.

The proposed NSGA-II procedure for our cell planning and joint cell and backhaul planning problem is shown in Algorithm 3.

\begin{algorithm}
\SetKwInOut{Input}{Input}\SetKwInOut{Output}{Output}
\caption{NSGA-II for planning problems}
\label{NSGAII}
\Input{$N_{\text{It}},N_{\text{BS}},f(\mu^{(.)}_{(i)}),\epsilon,\sigma_\mu,x_{\text{max}},x_{\text{min}},y_{\text{max}},y_{\text{min}},N_{\mathcal{F}}$}
\Output{$\mathbf{L}_{\text{W-BS}},\mathbf{L}_{\text{U-
BS}},\mathbf{L}_{\text{BS}},\mathcal{C}^b_f,\pmb{\mathcal{Z}}$,$N_{\text{BS}}$,$N_{\text{W-BS}}$,$N_{\text{U-BS}}$}
Generate an initial random population with size of $N$ for variables and compute $\mathbf{F}(\mathbf{x}_i),~\forall i\in N$.

Use \textbf{Algorithm 2} to rank the individuals, then calculate $d(\mathbf{x}_i),~ \forall i$.

\For{$i=1,2,\dots,N_{\text{It}}$}{Run \textbf{Algorithm 1}.

Sort solutions based on their crowding distances in descending order.

{\color{black}Run \textbf{Algorithm 2} to get $FR_k',~\forall k'$, sort solutions based on their fronts in ascending order.}

Retain the first $N$ individuals and ignore the others.}

\end{algorithm} 

\section{Results}
\subsection{System parameters for cell planning} 

We consider an area with mmWave BSs where some BSs have direct access to fiber (W-BSs) while the remaining ones are supported by W-BSs via wireless IBFD self-backhauling. The cell planning takes different criteria such as capacity, coverage, and self-backhauling into account. Specifically, parameters such as $N^{\text{cap}} _\text{BS}$, $N^ {\text{cov}} _\text{BS}$, and $\omega$ are set to default values, as shown in Table I. In addition, we assume that the height of antenna in UEs and BSs is set ot $h_{\text{UE}}=1.5$ m and $h_{\text{BS}}=2.5$ m, respectively. As listed in Table II, the required bandwidth for a data rate of $180$ Mbps is approximately $50$ MHz. On the other side, we also have the dimensions of areas, hereupon, by using \eqref{coverage} we can obtain $N^ {\text{cov}} _\text{BS}$. Accordingly, by utilizing the concluded values $N^ {\text{cov}} _\text{BS}$, $N^ {\text{cap}} _\text{BS}$, and $N_\text{BS}$ for each scenarios we can appropriately set up the optimization algorithms.

\begin{table}\label{CellPlanningTabel}
\caption{Simulation Parameters and Default Values}
\begin{center}
\begin{tabular}{|M{1.15cm}|M{1cm}||M{1.05cm}|M{1.8cm}|}
\hline
\multicolumn{4}{|c|}{\textbf{Cell Planning Parameters}} \\
\hline
\hline
\centering
\textbf{Parameter}&\textbf{Value}&\textbf{Parameter}&\textbf{Value}\\
\hline
$\beta^{\text{Access}}_{\text{los}}$ & $2$ &$\beta^{\text{Access}}_{\text{nlos}}$ & $3.3$\\
\hline
$\beta^{\text{Backhaul}}_{\text{los}}$ & $2$ &$\beta^{\text{Backhaul}}_{\text{nlos}}$ & $3.5$\\
\hline
$\sigma^{\text{Access}}_{\text{los}}$ & $5.2$ &$\sigma^{\text{Access}}_{\text{nlos}}$ & $7.2$\\
\hline
$\sigma^{\text{Backhaul}}_{\text{los}}$ & $4.2$ &$\sigma^{\text{Backhaul}}_{\text{nlos}}$ & $7.9$\\
\hline 
$\alpha_{\text{los}}$ & $70$ dB&$\alpha_{\text{nlos}}$ & $70$ dB\\
\hline
$N_{lim}$ & $3$ & $\tau$ & $10^{-5}$\\
\hline
$AG_{\text{BS}}^{a}$ & $18$ dBi&${AG}_{\text{BS}}^{b}$& $52$ dBi\\
\hline
${AG}_{\text{UE}}$ & $18$ dBi&$N_s$& $3$ \\
\hline
Noise Figure & $10$ dB& $\sigma^2_{N}$& $-174$ dBm/Hz\\
&&&+$10$$\log_{10}$(BW) \\
\hline
$\omega$ & $\frac{1}{3}$ & $BW_s$ & $4$ GHz \\
\hline 
$S_{\text{BS}}$& $\frac{3\sqrt{3}}{2} R_{\text{BS}}^2$& $BW_{RB}$ & $1$MHz\\
\hline
$\mathcal{P}_a$ & $1$ Watt&$\mathcal{P}_b$ & $1.26$ Watt\\
\hline
$\gamma^{th}_a$ & $10$ dB& $\gamma^{th}_b$ & $55$ dB\\
\hline
$\varrho^{th}_a$ & $0.9$&$\varrho^{th}_b$& $0.9$\\
\hline
$\delta_{cov}$& $0.8-0.9$ &$\delta_{cap}$& $0.8-0.9$\\
\hline 
\hline
\multicolumn{2}{|c||}{\textbf{Genetic Algorithm}}
&\multicolumn{2}{c|}{\textbf{Network Costs}}\\
\hline
\hline
\centering
\textbf{Parameter}&\textbf{Value}&\textbf{Facility}&\textbf{Price}\\
\hline
$N_{\text{pop}}$ & $100$ & $C_{U}$ & $1\$/\text{each}$\\
\hline
$\epsilon$ & $0.15$&$C_{W}$&$2\$/\text{each}$ \\
\hline
$\mu$ & $0.15$ & $C_f$ & $0.01\$/m$\\
\hline
$|\mathcal{N}_c|$&$0.9N_{pop}$&$C_d$ & $0.02\$/m$\\
\hline
$|\mathcal{N}_m|$&$0.4N_{pop}$&$C_s$ & $0.05\$/\text{each}$\\
\hline
\hline
\end{tabular}
\end{center}
\end{table}

\begin{figure}
\centering
\includegraphics[trim={4cm 8.5cm 4cm 8.5cm},width=7cm,clip]{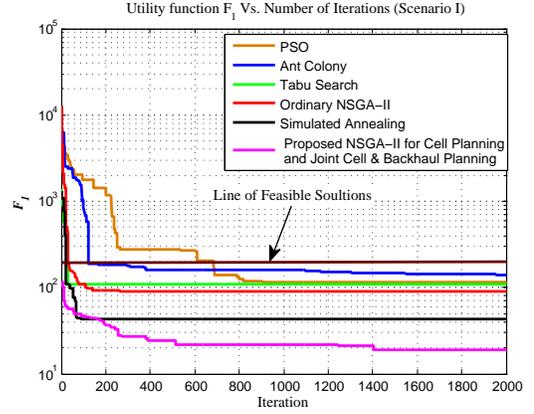}\\
\caption{Speed of convergence and final solution for $F_1$, Scenario I}\label{ConvergenceSpeed}
\end{figure}
\begin{figure}
\centering
\includegraphics[trim={4cm 8.5cm 4cm 8.5cm},width=7cm,clip]{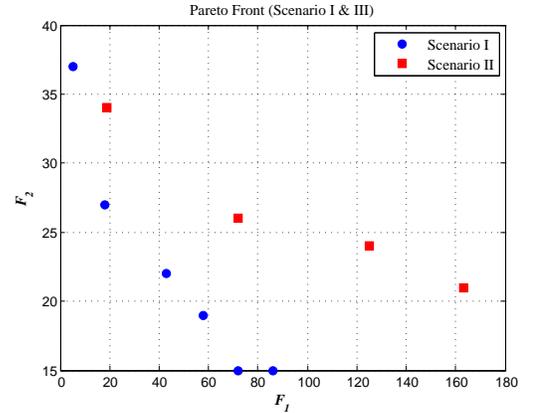}\\
\caption{Pareto fronts for Scenario I \& II}\label{ParetoAlgorithmI}
\end{figure}

{\color{black} Furthermore, the parameter settings for self-backhauled mmWave, equipment costs including both backhaul and access components, and parameter values of our proposed NSGA-II algorithm are listed in Table II. These parameter settings are taken from \cite{6834753}, \cite{rappaport2013millimeter},\cite{singh2015tractable}, \cite{6840343}, \cite{ranaweera2013design2}, and \cite{7499308}. We note that the NSGA-II parameters are empirically set such that the best results are achieved.}

\begin{table*}\label{Scenarios}
\caption{Area and Subarea Characteristics}
\begin{center}
\begin{tabular}{|M{1.2cm}||M{2.6cm}|M{2cm}|M{2cm}|M{1 cm}|M{1cm}|M{1cm}|}
\hline
\multicolumn{7}{|c|}{\textbf{Scenario Description}}\\
\hline
\hline
\hline
\centering
\textbf{Scenario}& $\mathbf{N^{\textbf{user}}_{s_i}}$&$\mathbf{S_{s_i}}$(Km$^2$)& $\!\!\pmb{\lambda_{\mathcal{S}_i}}$(user/Km$^2$)& $\!\mathbf{N}^{\textbf{Cov}}_{\textbf{BS}}$& $\!\mathbf{N}^{\textbf{Cap}}_{\textbf{BS}}$&$\!\mathbf{N_{Total}}$ \\
\hline
\hline

Scenario\newline I &$N^{\text{user}}_{s_1}=600\newline N^{\text{user}}_{s_2}=400$& $S_{s_1}=0.224$\newline$S_{s_2}=0.026$&$\lambda_{s_1}=2678.3$\newline$\lambda_{s_2}=15384$& $10$ & $5$ & $10$ \\
\hline
\hline

Scenario II &$N^{\text{user}}_{s_1}=400$ \newline $N^{\text{user}}_{s_2}=600$ \newline $N^{\text{user}}_{s_3}=800$ & $~S_{s_1}=0.224$\newline$~~S_{s_2}=0.026$\newline$~~S_{s_3}=0.026$&$\lambda_{s_1}=4464.2$\newline$~\lambda_{s_2}=38940$\newline$~\lambda_{s_3}=38940$& $10$ & $10$ & $10$ \\
\hline
\hline

Scenario III & $N^{\text{user}}_{s_1}=500\newline ~~N^{\text{user}}_{s_2}=1300\newline N^{\text{user}}_{s_3}=500\newline N^{\text{user}}_{s_4}=200$&$~S_{s_1}=0.25\newline ~~S_{s_2}=0.25\newline ~~S_{s_3}=0.25 \newline S_{s_4}=0.25$&$~\lambda_{s_1}=2000\newline~~\lambda_{s_2}=5200\newline~~\lambda_{s_3}=2000\newline \!\!\lambda_{s_4}=400$& $39$ & $13$ & $39$\\
\hline
\hline 
\end{tabular}
\end{center}
\end{table*}

\subsection{Scenarios}
We define some scenarios that consider different distributions of users as well as various environmental characteristics. In the first scenario, we consider a square area with two subareas. Users are distributed according to the Poisson point process (PPP) on both subareas. The second subarea consists of a square with diameter $d_{s_2} =100\sqrt{2}$ m in the middle of the area with different users distribution. Further details of the other considered scenarios are presented in Table III. As shown in the table, various scenarios are supposed  for the distributions of users and area dimensions to simulate a variety of possible deployment situations. Table II also specifies other important design parameters, including the total number of BSs for the capacity and coverage of the areas, which are derived from Table I and by using \eqref{numberofBSforCapacity} and \eqref{coverage} as well. We use a hexagonal cell to compute the area of each cell. Thus, the coverage area of each cell equals $S_{\text{BS}}=\frac{3\sqrt{3}}{2}R_{\text{BS}}^2$, where $R_{\text{BS}}$ denotes the cell radius.

\subsection{Performance Analysis of Cell Planning Algorithms}
Fig. {\ref{ConvergenceSpeed}} shows the speed of convergence and the final $F_1$ obtained under Scenario I. {\color{black} We also simulated other well-known algorithms such as an ordinary NSGA-II, particle swarm optimization (PSO), ant colony, tabu search (TS), and simulated annealing (SA). Comparing the results obtained from our proposed NSGA-II algorithm for {\color{black} cell planning} and joint cell and backhaul planning to the results of the aforementioned algorithms reveals that our method converges faster and achieves superior performance. Among the other algorithms, SA has better performance, though it remains in a local minimum, which our method is able to escape from. The superior performance of our proposed NSGA-II algorithm compared to the ordinary NSGA stems from the randomness introduced by $f(\mu)$ in Section IV.} The line of feasible realization indicates that all constraints are satisfied and the algorithms successfully search for the best solutions among the feasible set. We observed the same behavior for the other scenarios, however, due to the space limitations, we show only the results of Scenario I. {\color{black}
To compare the complexity of the proposed algorithm against the other methods, Table IV reports CPU running time for each of them. It should be noted that all tests were executed on a Windows 7-64 bit Professional with Intel Core i7 2.2 GHz CPU and 8GB RAM.}
\begin{table}\label{CellPlanningTabel}
\caption{CPU running time (second) }
\begin{center}
{\color{black}
\begin{tabular}{|M{2.4cm}|M{0.9 cm}|M{2.15cm}|M{0.9cm}|}

\hline
\centering
\textbf{Method}&\textbf{time}&\textbf{Method}&\textbf{time}\\
\hline
\hline
Proposed NSGA-II & $3678.96$ & Ordinary NSGA-II & $2621.98$\\
\hline
Tabu search& $8084.53$ & Ant Colony & $3396.29$\\
\hline
Simulated Annealing & $7700.47$ & PSO & $2438.39$\\
\hline
\end{tabular}
}
\end{center}
\end{table}

\begin{figure*}
    \centering
    \begin{subfigure}[t]{0.5\textwidth}
        \centering
        \includegraphics[trim={7.8cm 2.4cm 8.5cm 2.1cm},width=6.5cm,clip]{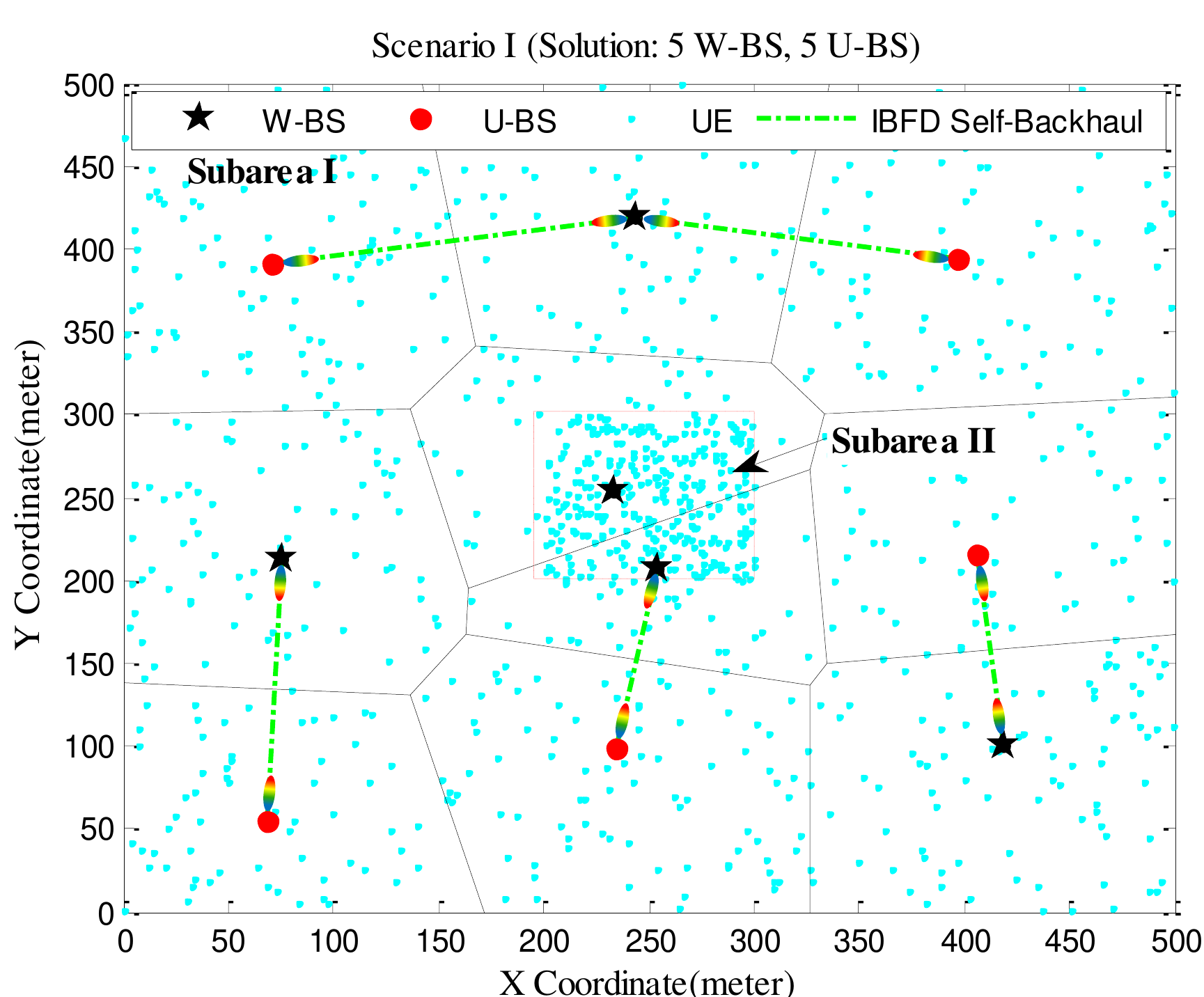}
        \caption{Realized solution with low cost}
        \label{CellPlanningScenarioIBestCost}
    \end{subfigure}%
    ~ 
    \begin{subfigure}[t]{0.5\textwidth}
        \centering
        \includegraphics[trim={7.8cm 2.4cm 8.3cm 2.1cm},width=6.5cm,clip]{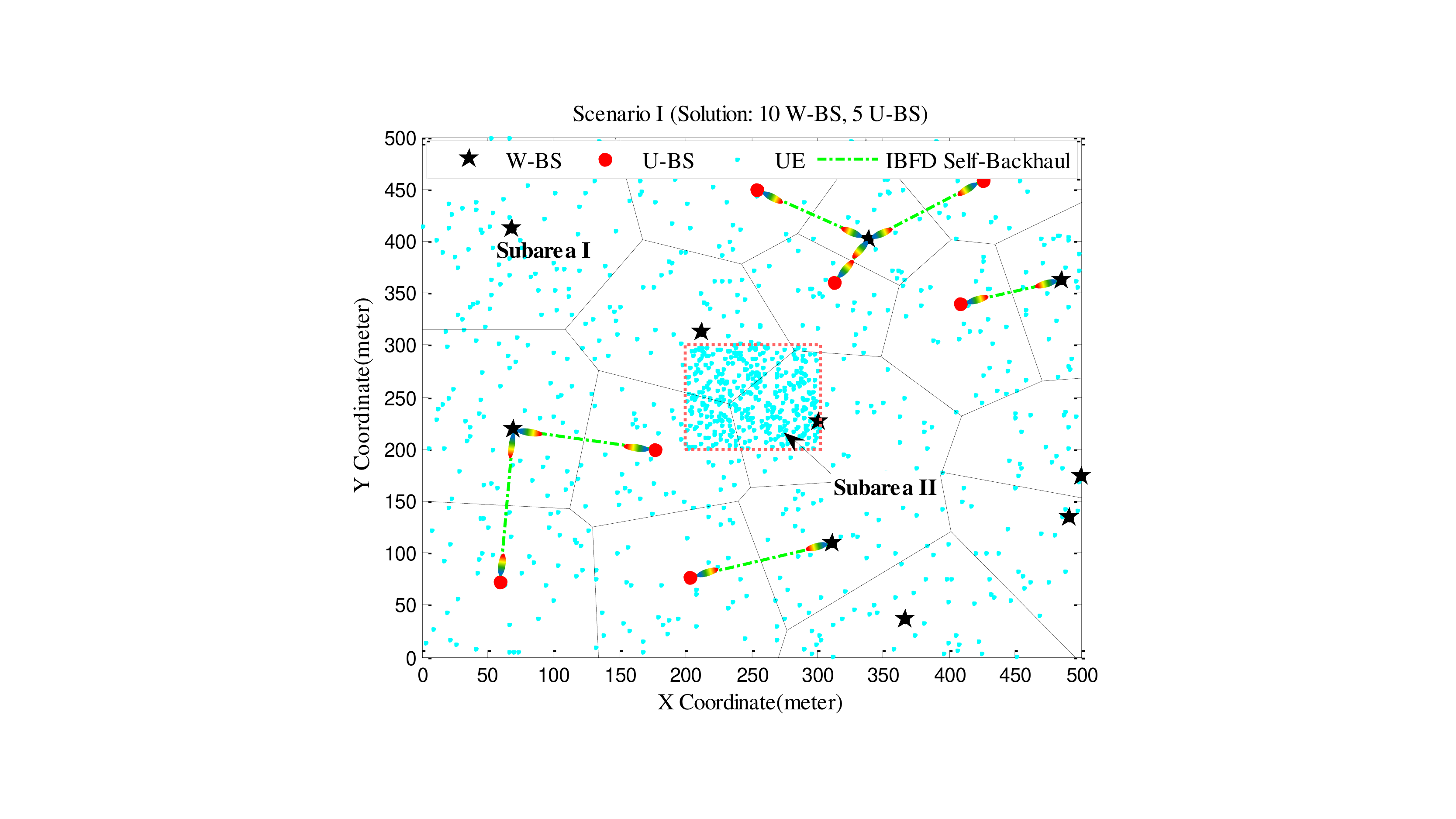}
        \caption{Realized solution with maximum coverage}
        \label{CellPlanningScenarioIBestCoverage}
    \end{subfigure}
    \caption{Cell layout with identifying IBFD self-backhauling links for Scenario I}
\end{figure*}

The Pareto front obtained for scenarios I and II are depicted in Fig. {\ref{ParetoAlgorithmI}}. We observe that the two objective functions of joint planing, $F_1$ and $F_2$, are opposite. This is due to the fact that more BSs should be deployed to serve UEs with better SINR and rate, therefore, this redundant components will increase the total cost.

The final cellular network realizations and topologies obtained from Algorithm I under Scenario I are illustrated in Fig. {\ref{CellPlanningScenarioIBestCost}} and Fig. {\ref{CellPlanningScenarioIBestCoverage}}, respectively. As it is clear, the realization with lowest cost has utilized less W-BS and U-BS, while on the other side, the realization with best coverage has more BSs. Additionally,  although the outputs are different, they have in common that both algorithms place W-BSs in congested regions. This is due to the fact that by deploying W-BSs in congested areas fewer resources are assigned to IBFD self-backhauling, whereby each W-BS can support at most $3$ U-BSs. As shown in both figures, W-BSs with a low number of associated users (referred to as self-load) support $3$ U-BSs, while W-BSs with high self-load support fewer or even no U-BSs. To give more insight about the complicated situation in Scenario II, We depicted one of the outlined cells' scheme in Fig. {\ref{ScenarioII}}.

\begin{figure*}
    \centering
    \begin{subfigure}[t]{0.5\textwidth}
        \centering
        \includegraphics[trim={4cm 8.5cm 4.5cm 8.5cm},width=6.5cm,clip]{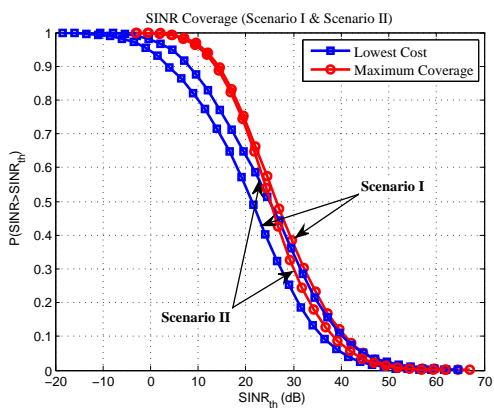}
        \caption{SINR coverage for two extreme cases.}
        \label{SINRScenarioI&II}
    \end{subfigure}%
    ~ 
    \begin{subfigure}[t]{0.5\textwidth}
        \centering
        \includegraphics[trim={4cm 8.5cm 4.5cm 8.5cm},width=6.5cm,clip]{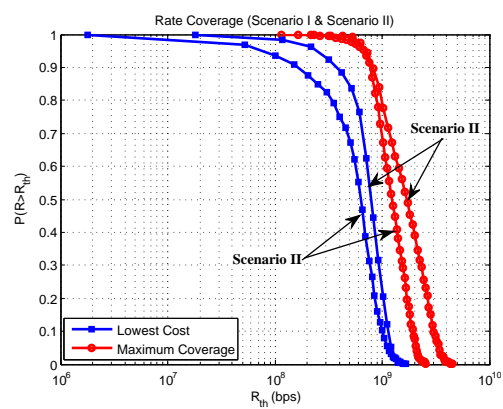}
        \caption{Rate coverage for two extreme cases.}
        \label{RateScenarioI&II}
    \end{subfigure}
    \caption{Comparison of SINR and Rate coverage under different scenarios and cases.}
\end{figure*}

For further insights into the performance of our proposed algorithms,  we investigate the SINR and rate coverage for Scenarios I and II. Fig. {\ref{SINRScenarioI&II}} depicts the SINR coverage. In this figures, we observe that the realization with best coverage (maximum coverage in the figure) has offered better SINR coverage due to utilizing more BSs, whereas, the solution with lowest cost has worse SINR coverage. By comparing the SINR coverage under different scenarios, we also observe that this gap varies with respect to the densities of subareas in the scenarios.

The rate coverage under scenarios I and II is shown in Fig. {\ref{RateScenarioI&II}}. We observe that most users achieve a data rate of $200$ Mbps. The same as before, we demonstrate two realized solutions, one with maximum coverage and higher cost, and the other with lowest coverage and lower cost.

\subsection{mmWave Cell Planning with Blockage Effect}
As mmWave cellular networks suffer from the blockage of obstacles, in this subsection we investigate the proposed mmWave cell planning by considering the blockage effect. Toward this aim, we define binary variable $\Pi_{i}^{j}$ which is 1 if the direct link between pixels $i$ and $j$ is blocked by an obstacle, otherwise it is 0. Furthermore, we modify ${\mathcal{X}^p_i}$ as follows

\begin{equation}\label{PixelAssociation2}
{\mathcal{X'}^i_p}=\begin{cases}1,& \text{if}\; i=\argmin\limits_{k\in\mathcal{B}} \frac{||\pmb{\mathscr{L}}_p-\pmb{\mathscr{L}}_{k}||}{\Pi_{i}^{p}}, \\ 0, & \qquad \qquad \; oth.
\end{cases}
,\; \forall p\in \mathcal{P}.
\end{equation}
In addition, the same formula is applied to modify ${\mathcal{Y'}_u^w}$. It should be noted that if for a pixel such as $p$, we have ${\mathcal{X'}_p^i}=\infty, \forall i\in\mathcal{B}$ that pixel will be excluded from the optimization procedure (it means that pixel {\color{black} $p$ is surrounded by obstacles}, and it is impossible to cover it). However, this approach is not employed for ${\mathcal{Y'}_u^w}$. 

Finally, in order to model the blockage effect, we substitute ${\mathcal{X'}_p^i}$ and ${\mathcal{Y'}_u^w}$ in the proposed formulations in subsection III.B. By accounting  obstacles into the cell planning problem, it will become more complicated, however, thanks to the Bresenham's line algorithm \cite{5388473}, its computational complexity can be mitigated efficiently.

Fig. {\ref{BlockageModeling}} indicates the results of cell planning over an area with four rectangular obstacles in which users are distributed on the whole area randomly. The figure shows the outline of cells and their corresponding IBFD self-backhauling links obtained from a point on its corresponding Pareto front. As the result shows, the positions of W-BSs as well as U-BSs are adapted according to the location of blocks. It means that they are placed where to minimize the blocks' effect. Successively, their locations are where that no obstacle disrupts IBFD self-backhauling links.
\begin{figure}
\centering
\includegraphics[trim={7.8cm 2.4cm 8.5cm 2.1cm},width=7cm,clip]{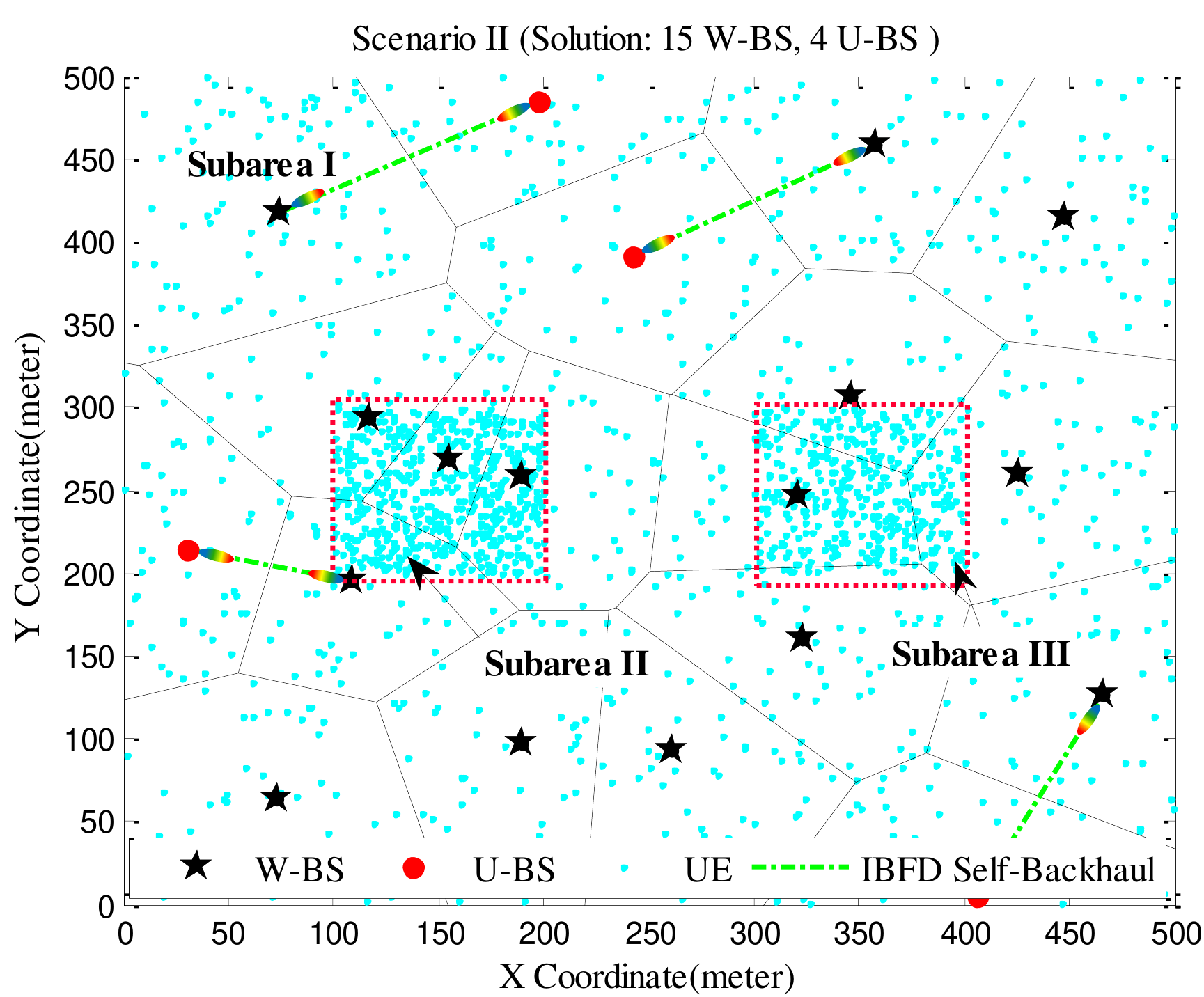}\\
\caption{One realized cell's outline for Scenario II }\label{ScenarioII}
\end{figure}
\begin{figure}
\centering
\includegraphics[trim={3.75cm 8.5cm 4.7cm 8.8cm},width=7cm,clip]{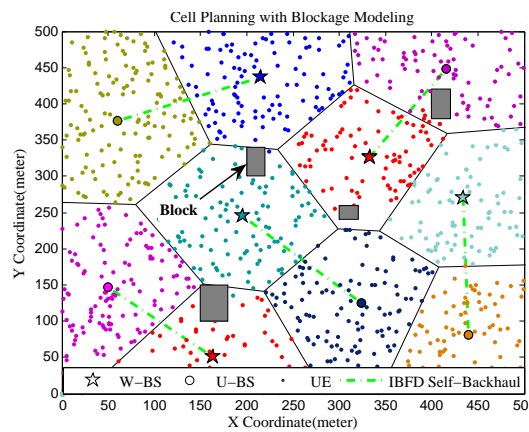}\\
\caption{Cell's outline with blockage modeling.}\label{BlockageModeling}
\end{figure}

\begin{figure}
\centering
\includegraphics[trim={3.75cm 8.5cm 4.7cm 8.8cm},width=7cm,clip]{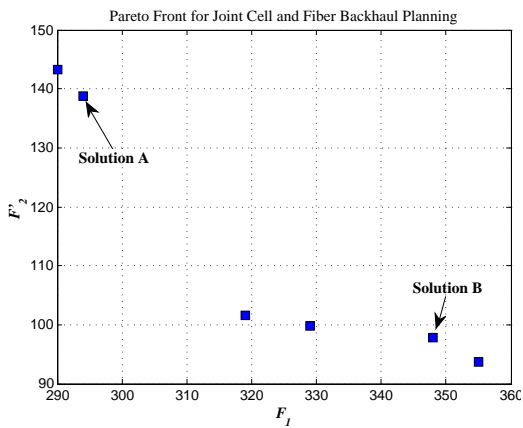}\\
\caption{Pareto front of NGSA-II.}\label{ParetoFour}
\end{figure}

\begin{figure*}[t]
    \centering
    \begin{subfigure}[t]{0.5\textwidth}
        \centering
        \includegraphics[trim={3.8cm 8.6cm 4cm 8.5cm},width=7cm,clip]{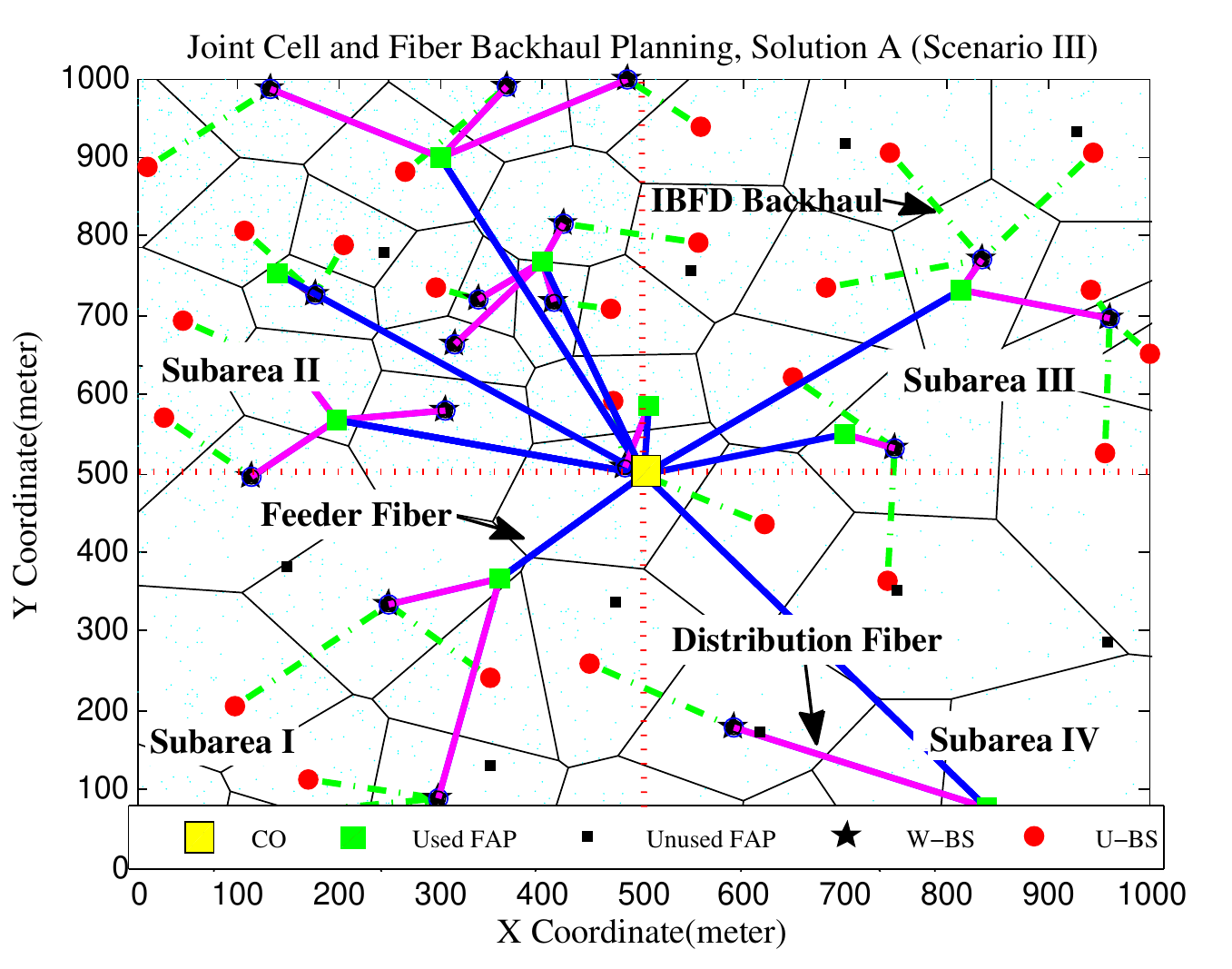}
        \caption{First Pareto point.}
        \label{CellFiberPlanningWirelessPoint}
    \end{subfigure}%
    ~ 
    \begin{subfigure}[t]{0.5\textwidth}
        \centering
        \includegraphics[trim={3.8cm 8.6cm 4cm 8.4cm},width=7cm,clip]{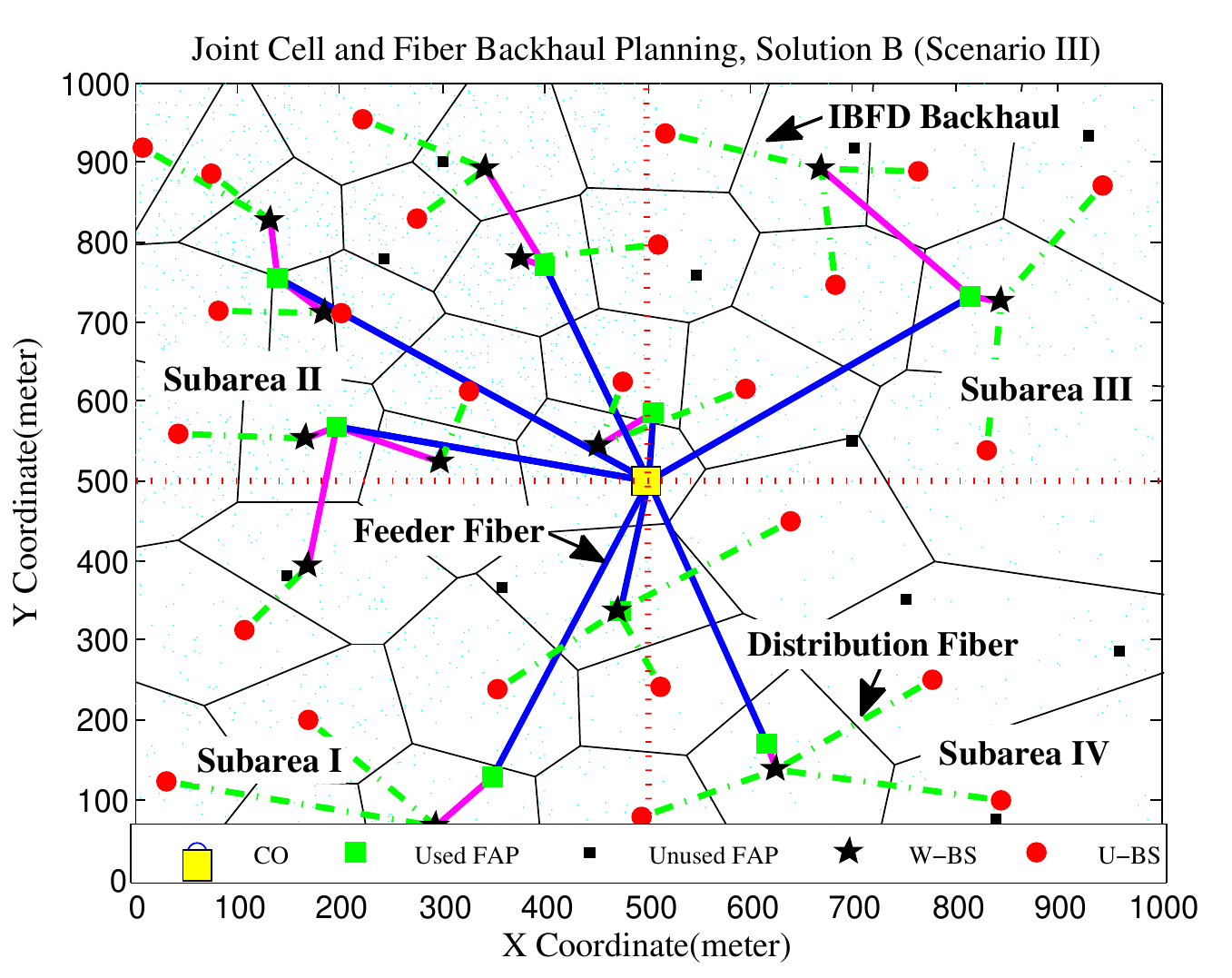}
        \caption{Second Pareto point.}
    	\label{CellFiberPlanningFiberPoint}
    \end{subfigure}
    \caption{Cell and fiber backhaul layout under Scenario IV.}
\end{figure*}

\subsection{Performance Analysis of Joint Cell and Fiber Backhaul Planning}
Recall that we leverage existing dark feeder fibers and FAPs to support mmWave cellular network via a fiber backhaul. Although existing dark fibers may be used as distribution fibers, in this study, we assume that new distribution fibers are deployed to connect a W-BS to its corresponding FAP. In all scenarios, a finite number of FAPs are distributed randomly across the areas under considerations. Moreover, we assume that the central office is located at the middle of the areas. The cost of the components needed to realize the fiber backhaul infrastructure (including dark feeder fibers) is given in Table I,  we also assume that $\mathcal{R}=4$. It should be noted that carriers can easily replace their real data including available FAPs, prices, and CO locations.   

We considered the same scenarios as in the previous subsection. However, due to to space limitations, in the following we only report the results of Scenarios III. We assume that and $20$ FAPs are distributed randomly in the areas under Scenario III. The second objective is to minimize the total cost which includes both cell and fiber backhaul deployments expenses, denoted by $F'_2$. We again apply the NSGA-II algorithm because of its proficiency in solving multi-objective optimization problems. The initial populations for NSGA-II is set to $200$ and the total number of iterations equals $2000$.

The Pareto front obtained from the NSGA-II algorithm is depicted in Fig. {\ref{ParetoFour}}. The same as previous section, we can observe that the two objective functions of the joint planing, $F_1$ and $F_2$, are opposite. This is due to the fact that more fibers must be deployed to serve W-BSs when W-BSs are located further away from FAPs in order to serve more UEs and distributed U-BSs, which resulting in a higher cost function. This is in addition to utilizing more BSs as well.

For illustration, two solutions of the obtained Pareto front (Solution A and Solution B) are selected and investigated. The point A has better coverage but at the expense of higher deployment costs. The second one, point B, has worse coverage but with less deployment costs. In Figs. {\ref{CellFiberPlanningWirelessPoint}} and {\ref{CellFiberPlanningFiberPoint}} the final topologies of the selected points are shown under Scenario III. We observe that in the topology of the first Pareto point (the solution with the lowest fiber deployment costs) W-BSs are located near FAPs to use less distribution fiber, whereas in the second Pareto point (the solution with the best LOS coverage), W-BSs are far away from FAPs to have a better LOS coverage meanwhile more BSs are also utilized, at the expense of higher implementation costs.  

\begin{figure}
\centering
\includegraphics[trim={4cm 8.5cm 4cm 8.5cm},width=7cm,clip]{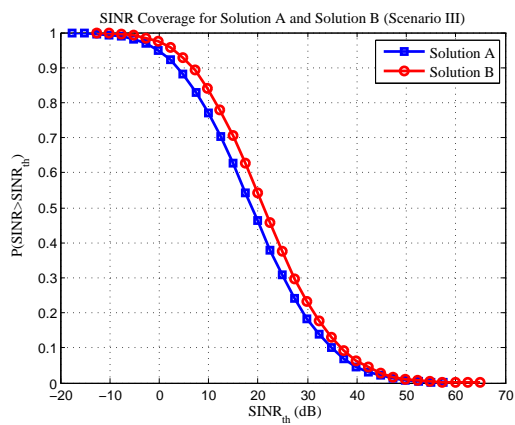}\\
\caption{SINR coverage under scenarios I and IV for the selected Pareto points.}
\label{CellFiberPlanningSNR}
\end{figure}

\begin{figure}
\centering
\includegraphics[trim={4cm 8.5cm 4cm 8.5cm},width=7cm,clip]{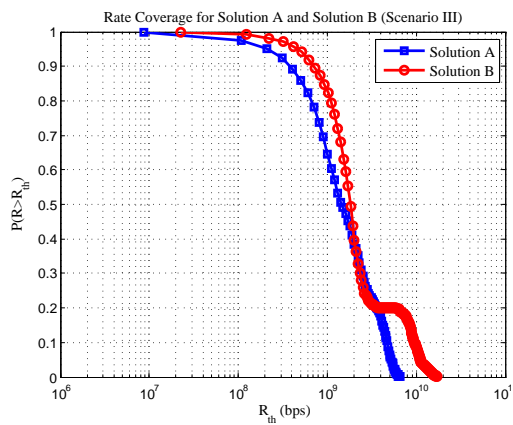}\\
\caption{Rate coverage under scenarios I and IV for the selected Pareto points.}
\label{CellFiberPlanningRate}
\end{figure}

Fig. {\ref{CellFiberPlanningSNR}} and Fig. {\ref{CellFiberPlanningRate}} indicate the SINR and rate coverage of the selected points under Scenarios III, respectively. In realistic situations based on the policies of cellular operators, a trade-off  between deployment costs and SINR/Rate coverage has to be made by selecting the desired solution from the obtained Pareto front. 
 
\begin{figure}
\centering
\includegraphics[trim={4cm 8.5cm 4cm 8.5cm},width=7cm,clip]{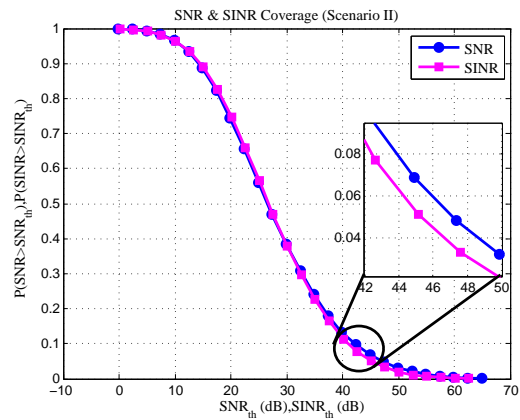}\\
\caption{SNR and SINR Coverage (Scenario III)}
\label{SNR&SINR}
\end{figure}

We also investigate whether we can neglect the effect of interference in the proposed scenarios or not. To do so, let assume the extreme case, Scenario III, in which we have the highest users density, and therefore, more BSs are utilized. In this case, the SNR and SINR coverage of Scenario III are compared in Fig. {\ref{SNR&SINR}}. We can observe that the SNR and SINR coverage are very close to each other, then, we can claim the interference in such a dense area is low enough to be ignored. Therefore, we can approximate SINR with SNR which brings less computational complexity. This effect of interference is studied well in \cite{7499308}, where the authors have determined the regime of noise-limited and interference-limited performances with respect to the density of users and BSs.

\section{Conclusions}
We studied a cell planning problem for a two-tier cellular network to determine the number and the location of BSs with fiber backhaul (W-BSs) and BSs with wireless self-backhauled (U-BSs). We proposed an algorithm to minimize deployment costs while satisfying given cell and capacity coverage constraints. Furthermore, we applied a well-known meta-heuristic algorithm to solve our proposed cell planning problem.

{\color {black} The proposed planning programs were formulated as a multi-objective optimization problem by including the cell and fiber backhaul deployment at lowest costs. Further, we optimally developed an efficient meta-heuristic algorithm which has superiority against other well-known algorithms.}   

In order to {\color{black} scrutinize} the performance of the proposed algorithms, we considered three different deployment scenarios with different user spatial distributions and coverage areas. {\color{black}With both planning problems}, W-BSs are placed in congested areas with higher user density to consume fewer resources for IBFD self-backhauled links. Furthermore, {\color{black}our results on both cell planning} and joint cell and fiber backhaul planning show that there exists a trade-off between deployment costs and SINR/rate coverage. Based on given cellular network operator polices the desired solution can be selected from the Pareto front of our studied multi-objective problem.

\bibliographystyle{IEEEtran}

\bibliography{IEEEabrv}

\begin{thebibliography}{10}
\providecommand{\url}[1]{#1}
\csname url@samestyle\endcsname
\providecommand{\newblock}{\relax}
\providecommand{\bibinfo}[2]{#2}
\providecommand{\BIBentrySTDinterwordspacing}{\spaceskip=0pt\relax}
\providecommand{\BIBentryALTinterwordstretchfactor}{4}
\providecommand{\BIBentryALTinterwordspacing}{\spaceskip=\fontdimen2\font plus
\BIBentryALTinterwordstretchfactor\fontdimen3\font minus
  \fontdimen4\font\relax}
\providecommand{\BIBforeignlanguage}[2]{{%
\expandafter\ifx\csname l@#1\endcsname\relax
\typeout{** WARNING: IEEEtran.bst: No hyphenation pattern has been}%
\typeout{** loaded for the language `#1'. Using the pattern for}%
\typeout{** the default language instead.}%
\else
\language=\csname l@#1\endcsname
\fi
#2}}
\providecommand{\BIBdecl}{\relax}
\BIBdecl

\bibitem{Cisco}
Cisco, ``Cisco visual index: Global mobile data traffic forecast update,
  2015-2020.'' Whitepaper, available at: http://goo.gl/SwuEIc.

\bibitem{6824752}
J.~G. Andrews \emph{et~al.}, ``What will 5{G} be?'' \emph{IEEE J. Sel. Areas
  Commun.}, vol.~32, pp. 1065--1082, June 2014.

\bibitem{6736750}
W.~Roh \emph{et~al.}, ``Millimeter-wave beamforming as an enabling technology
  for 5{G} cellular communications: theoretical feasibility and prototype
  results,'' \emph{IEEE Commun. Mag.}, vol.~52, pp. 106--113, Feb. 2014.

\bibitem{6824746}
A.~Ghosh \emph{et~al.}, ``Millimeter-wave enhanced local area systems: A
  high-data-rate approach for future wireless networks,'' \emph{IEEE J. Sel.
  Areas Commun.}, vol.~32, pp. 1152--1163, June 2014.

\bibitem{6834753}
M.~R. Akdeniz \emph{et~al.}, ``Millimeter wave channel modeling and cellular
  capacity evaluation,'' \emph{IEEE J. Sel. Areas Commun.}, vol.~32, pp.
  1164--1179, June 2014.

\bibitem{6963798}
X.~Ge, H.~Cheng, M.~Guizani, and T.~Han, ``5{G} wireless backhaul networks:
  challenges and research advances,'' \emph{IEEE Network}, vol.~28, pp. 6--11,
  Nov. 2014.

\bibitem{6588652}
C.~Ranaweera \emph{et~al.}, ``Design and optimization of fiber optic small-cell
  backhaul based on an existing fiber-to-the-node residential access network,''
  \emph{IEEE Commun. Mag}, vol.~51, pp. 62--69, Sept. 2013.

\bibitem{7218503}
H.~Beyranvand \emph{et~al.}, ``{FiWi} enhanced {LTE-A} {HetNets} with
  unreliable fiber backhaul sharing and {WiFi} offloading,'' in \emph{2015 IEEE
  Conference on Computer Communications (INFOCOM)}, Apr. 2015, pp. 1275--1283.

\bibitem{7306543}
H.~Dahrouj \emph{et~al.}, ``Cost-effective hybrid {RF/FSO} backhaul solution
  for next generation wireless systems,'' \emph{IEEE Wireless Communications},
  vol.~22, pp. 98--104, Oct. 2015.

\bibitem{7115912}
Y.~Li \emph{et~al.}, ``Optimization of free space optical wireless network for
  cellular backhauling,'' \emph{IEEE J. Sel. Areas Commun.}, vol.~33, pp.
  1841--1854, Sept. 2015.

\bibitem{7947084}
Y.~Yu \emph{et~al.}, ``Hybrid fiber-wireless network: an optimization framework
  for survivable deployment,'' \emph{IEEE/OSA Journal of Optical Communications
  and Networking}, vol.~9, no.~6, pp. 466--478, June 2017.

\bibitem{7925837}
M.~N. Islam, S.~Subramanian, and A.~Sampath, ``Integrated access backhaul in
  millimeter wave networks,'' in \emph{2017 IEEE Wireless Communications and
  Networking Conference (WCNC)}, March 2017, pp. 1--6.

\bibitem{Interdigital}
Cisco, ``Small cell millimeter wave mesh backhaul,'' Feb. 2013. Whitepaper,
  available at: http://goo.gl/Dl2Z6V.

\bibitem{bharadia2013full}
D.~Bharadia, E.~McMilin, and S.~Katti, ``Full duplex radios,'' \emph{ACM
  SIGCOMM Computer Communication Review}, vol.~43, no.~4, pp. 375--386, 2013.

\bibitem{7306541}
R.~A. Pitaval \emph{et~al.}, ``Full-duplex self-backhauling for small-cell 5{G}
  networks,'' \emph{IEEE Wireless Communications}, vol.~22, pp. 83--89, Oct.
  2015.

\bibitem{1230131}
E.~Amaldi \emph{et~al.}, ``Planning {UMTS} base station location: optimization
  models with power control and algorithms,'' \emph{IEEE Trans. Wireless
  Commun.}, vol.~2, pp. 939--952, Sept. 2003.

\bibitem{6196268}
G.~Koutitas, A.~Karousos, and L.~Tassiulas, ``Deployment strategies and energy
  efficiency of cellular networks,'' \emph{IEEE Trans. Wireless Commun.},
  vol.~11, pp. 2552--2563, July 2012.

\bibitem{7450686}
W.~Zhao, S.~Wang, C.~Wang, and X.~Wu, ``Approximation algorithms for cell
  planning in heterogeneous networks,'' \emph{IEEE Trans. Veh. Technol.},
  vol.~PP, pp. 1--1, 2016.

\bibitem{7056465}
H.~Ghazzai \emph{et~al.}, ``Optimized {LTE} cell planning with varying spatial
  and temporal user densities,'' \emph{IEEE Transactions on Vehicular
  Technology}, vol.~65, pp. 1575--1589, Mar. 2016.

\bibitem{7883847}
A.~Taufique \emph{et~al.}, ``Planning wireless cellular networks of future:
  Outlook, challenges and opportunities,'' \emph{IEEE Access}, vol.~5, pp.
  4821--4845, 2017.

\bibitem{7305743}
X.~Xu, W.~Saad, X.~Zhang, X.~Xu, and S.~Zhou, ``Joint deployment of small cells
  and wireless backhaul links in next-generation networks,'' \emph{IEEE Commun.
  Letters}, vol.~19, pp. 2250--2253, Dec. 2015.

\bibitem{7450185}
S.~S. Szyszkowicz \emph{et~al.}, ``Automated placement of individual
  millimeter-wave wall-mounted base stations for line-of-sight coverage of
  outdoor urban areas,'' \emph{IEEE Wireless Commun. Letters}, vol.~5, pp.
  316--319, June 2016.

\bibitem{7386685}
N.~Wang, E.~Hossain, and V.~K. Bhargava, ``Joint downlink cell association and
  bandwidth allocation for wireless backhauling in two-tier hetnets with
  large-scale antenna arrays,'' \emph{IEEE Transactions on Wireless
  Communications}, vol.~15, pp. 3251--3268, May 2016.

\bibitem{7505948}
T.~M. Nguyen, A.~Yadav, W.~Ajib, and C.~Assi, ``Resource allocation in two-tier
  wireless backhaul heterogeneous networks,'' \emph{IEEE Transactions on
  Wireless Communications}, vol.~15, pp. 6690--6704, Oct 2016.

\bibitem{5Gstandard1}
M.~Shafi, A.~F. Molisch, P.~J. Smith, T.~Haustein, P.~Zhu, P.~D. Silva,
  F.~Tufvesson, A.~Benjebbour, and G.~Wunder, ``5g: A tutorial overview of
  standards, trials, challenges, deployment, and practice,'' \emph{IEEE Journal
  on Selected Areas in Communications}, vol.~35, no.~6, pp. 1201--1221, June
  2017.

\bibitem{5Gstandard2}
S.~Parkvall, E.~Dahlman, A.~Furuskar, and M.~Frenne, ``Nr: The new 5g radio
  access technology,'' \emph{IEEE Communications Standards Magazine}, vol.~1,
  no.~4, pp. 24--30, Dec 2017.

\bibitem{jain2011practical}
M.~Jain~\textit{et al.}, ``Practical, real-time, full duplex wireless,'' in
  \emph{Proceedings of the 17th annual international conference on Mobile
  computing and networking}.\hskip 1em plus 0.5em minus 0.4em\relax ACM, Sept.
  2011, pp. 301--312.

\bibitem{7817893}
A.~Sharma, R.~K. Ganti, and J.~K. Milleth, ``Joint backhaul-access analysis of
  full duplex self-backhauling heterogeneous networks,'' \emph{IEEE Trans.
  Wireless Commun.}, vol.~16, pp. 1727--1740, March 2017.

\bibitem{6832464}
A.~Sabharwal \emph{et~al.}, ``In-band full-duplex wireless: Challenges and
  opportunities,'' \emph{IEEE J. Sel. Areas Commun.}, vol.~32, pp. 1637--1652,
  Sept. 2014.

\bibitem{6289432}
Y.~Luo \emph{et~al.}, ``Time- and wavelength-division multiplexed passive
  optical network {(TWDM-PON)} for next-generation {PON} stage 2 ({NG-PON2}),''
  \emph{J. Lightw. Technol.}, vol.~31, pp. 587--593, Feb. 2013.

\bibitem{ITU}
``40-gigabit-capable passive optical networks ({NG-PON2}): Physical media
  dependent (pmd) layer specification,,'' ITUT G.989.2 Recommendation, 2014.

\bibitem{7389581}
J.~S. Wey \emph{et~al.}, ``Physical layer aspects of {NG-PON2} standards
  2014;part 1: Optical link design [invited],'' \emph{IEEE/OSA Journal of
  Optical Communications and Networking}, vol.~8, pp. 33--42, Jan. 2016.

\bibitem{rappaport2013millimeter}
T.~S. Rappaport~\textit{et al}., ``Millimeter wave mobile communications for
  5{G} cellular: It will work!'' \emph{Access, IEEE}, vol.~1, pp. 335--349, May
  2013.

\bibitem{singh2015tractable}
S.~Singh~\textit{et al}., ``Tractable model for rate in self-backhauled
  millimeter wave cellular networks,'' \emph{IEEE J. Sel. Areas Commun.},
  vol.~33, pp. 2196--2211, Oct. 2015.

\bibitem{6840343}
T.~Bai, R.~Vaze, and R.~W. Heath, ``Analysis of blockage effects on urban
  cellular networks,'' \emph{IEEE Trans. Commun.}, vol.~13, pp. 5070--5083,
  Sept. 2014.

\bibitem{6994333}
D.~Kreutz \emph{et~al.}, ``Software-defined networking: A comprehensive
  survey,'' \emph{Proceedings of the IEEE}, vol. 103, pp. 14--76, Jan. 2015.

\bibitem{6898939}
P.~Rost \emph{et~al.}, ``Cloud technologies for flexible 5{G} radio access
  networks,'' \emph{IEEE Commun. Mag.}, vol.~52, pp. 68--76, May 2014.

\bibitem{drezner2001facility}
Z.~Drezner and H.~W. Hamacher, \emph{Facility location: applications and
  theory}.\hskip 1em plus 0.5em minus 0.4em\relax Springer Science \& Business
  Media, 2001.

\bibitem{996017}
K.~Deb, A.~Pratap, S.~Agarwal, and T.~Meyarivan, ``A fast and elitist
  multiobjective genetic algorithm: {NSGA-II},'' \emph{IEEE Trans. Evol.
  Comput.}, vol.~6, pp. 182--197, Apr. 2002.

\bibitem{srinivas1994muiltiobjective}
N.~Srinivas and K.~Deb, ``Muiltiobjective optimization using nondominated
  sorting in genetic algorithms,'' \emph{Evolutionary computation}, vol.~2, pp.
  221--248, 1994.

\bibitem{ranaweera2013design2}
C.~S. Ranaweera~\textit{et al.}, ``Design of cost-optimal passive optical
  networks for small cell backhaul using installed fibers,'' \emph{Journal of
  Optical Communications and Networking}, vol.~5, pp. A230--A239, Oct. 2013.

\bibitem{7499308}
M.~Rebato \emph{et~al.}, ``Understanding noise and interference regimes in 5g
  millimeter-wave cellular networks,'' in \emph{European Wireless 2016; 22th
  European Wireless Conference}, May 2016.

\bibitem{5388473}
J.~E. Bresenham, ``Algorithm for computer control of a digital plotter,''
  \emph{IBM Systems Journal}, vol.~4, no.~1, pp. 25--30, 1965.

\end{thebibliography}

\end{document}